\documentclass{emulateapj}
\usepackage{apjfonts}

\newcommand{\SPSD}{A}
\newcommand{\SIGR}{\sigma_r^2}

  \submitted{Accepted for publication in The Astrophysical Journal}
 
\begin{document}

\bibliographystyle{apj}

\title{Parameter Estimation from Time-Series Data with Correlated Errors: \\
	A Wavelet-Based Method and its Application to Transit Light Curves}

\author{
Joshua A.~Carter and
Joshua N.~Winn
}


\affil{Department of Physics, and Kavli Institute for
  Astrophysics and Space Research, \\Massachusetts Institute of
  Technology, Cambridge, MA 02139}

\email{carterja@mit.edu; jwinn@mit.edu}

\begin{abstract}

  We consider the problem of fitting a parametric model to time-series
  data that are afflicted by correlated noise. The noise is
  represented by a sum of two stationary Gaussian processes: one that
  is uncorrelated in time, and another that has a power spectral
  density varying as $1/f^\gamma$. We present an accurate and fast
  [$O(N)$] algorithm for parameter estimation based on computing the
  likelihood in a wavelet basis. The method is illustrated and tested
  using simulated time-series photometry of exoplanetary transits,
  with particular attention to estimating the midtransit time. We
  compare our method to two other methods that have been used in the
  literature, the time-averaging method and the residual-permutation
  method. For noise processes that obey our assumptions, the algorithm
  presented here gives more accurate results for midtransit times and
  truer estimates of their uncertainties.
 
\end{abstract}

  
\section{Introduction}
 
Frequently one wishes to fit a parametric model to time-series data
and determine accurate values of the parameters and reliable estimates
for the uncertainties in those parameters. It is important to gain a
thorough understanding of the noise and develop appropriate methods
for parameter estimation, especially at the research frontier, where
the most interesting effects are often on the edge of
detectability. Underestimating the errors leads to unjustified
confidence in new results, or confusion over apparent contradictions
between different data sets. Overestimating the errors inhibits
potentially important discoveries.

When the errors in the data are well understood and uncorrelated, the
problem of parameter estimation is relatively straightforward (see,
e.g., Bevington \& Robinson 2003, Gould 2003, Press et
al.~2007). However, when the noise is not well-understood---and
particularly when the noise exhibits correlations in time---the
problem is more challenging (see, e.g., Koen \& Lombard 1993, Beran
1994). Traditional methods that ignore correlations often give
parameter estimates that are inaccurate and parameter errors that are
underestimated. Straightforward generalization of the traditional
methods is computationally intensive, with time-complexity $O(N^2)$ in
the worst cases (where $N$ is the number of data points). This makes
certain analyses impractical.

Our specific concern in this paper is the analysis of time-series
photometry of exoplanetary transits. During a transit, a planet passes
in front of the disk of its parent star, which is evident from the
slight diminution in the light received from the star. A model of a
transit light curve may have many parameters, but we focus mainly on a
single parameter, the midtransit time $t_c$, for three reasons. The
first reason is the simplicity of a single-parameter model. The second
reason is that $t_c$ is a unique piece of information regarding each
transit event, and as such, the accuracy cannot be improved by
combining results from multiple transit observations. Instead one must
make the most of single-event observations even if they are afflicted
by correlated noise. The third reason is that transit timing offers a
means of discovering additional planets or satellites by seeking
anomalies in a sequence of transit times due to gravitational
perturbations [Holman \& Murray (2005), Agol et
al.~(2005)].\footnote{The transit duration is also expected to vary in
  the presence of additional gravitating bodies; see, e.g., Kipping
  (2009).}

Beginning with the work of Pont, Zucker, \& Queloz (2006), it has been
widely recognized that time-correlated noise (``red noise'') is a
limiting factor in the analysis of transit light curves. Many
practitioners have attempted to account for correlated errors in their
parameter estimation algorithms (see, e.g., Bakos et al.~2006, Gillon
et al.\ 2006; Winn et al.~2007, 2009; Southworth 2008). Among these
schemes are the ``time-averaging'' method, in which the effects of
correlations are assessed by computing the scatter in a time-binned
version of the data (Pont et al.~2006) and the
``residual-permutation'' method, a variant of bootstrap analysis that
preserves the time ordering of the residuals (Jenkins et al.~2002).

In this paper we present an alternative method for parameter
estimation in the presence of time-correlated noise, and compare it to
those two previously advocated methods. The method advocated here is
applicable to situations in which the noise is well described as the
superposition of two stationary (time-invariant) Gaussian noise
processes: one which is uncorrelated, and the other of which has a
power spectral density varying as $1/f^\gamma$.

A more traditional approach to time-correlated noise is the framework
of autoregressive moving average (ARMA) processes (see, e.g., Box \&
Jenkins 1976). The ARMA noise models can be understood as
complementary to our $1/f^\gamma$ model, in that ARMA models are
specified in the time domain as opposed to the frequency domain, and
they are most naturally suited for modeling short-range correlations
(``short-memory'' processes) as opposed to long-range correlations
(``long-memory'' processes). Parameter estimation with ARMA models in
an astronomical context has been discussed by Koen \& Lombard (1993),
Konig \& Timmer (1997), and Timmer et al.~(2000). As we will explain,
our method accelerates the parameter estimation problem by taking
advantage of the discrete wavelet transform. It is based on the fact
that a the covariance matrix of a $1/f^\gamma$ noise process is nearly
diagonal in a wavelet basis. As long as the actual noise is reasonably
well described by such a power law, our method is attractive for its
simplicity, computational speed, and ease of implementation, in
addition to its grounding in the recent literature on signal
processing.

The use of the wavelets in signal processing is widespread, especially
for the restoration, compression, and denoising of images (see, e.g.,
Mallat 1999). Parameter estimation using wavelets has been considered
but usually for the purpose of estimating {\it noise} parameters
(Wornell 1996). An application of wavelets to the problem of linear
regression with correlated noise was given by Fadili \& Bullmore
(2002). What is new in this work is the extension to an arbitary
nonlinear model, and the application to transit light curves.

This paper is organized as follows. In \S~\ref{sec:noise}, we review
the problem of estimating model parameters from data corrupted by
noise, and we review some relevant noise models. In
\S~\ref{sec:wavelet} we present the wavelet method and those aspects
of wavelet theory that are needed to understand the method. In
\S~\ref{sec:analysis}, we test the method using simulated transit
light curves, and compare the results to those obtained using the
methods mentioned previously. In \S~\ref{sec:conc} we summarize the
method and the results of our tests, and suggest some possible
applications and extensions of this work.
  
\section{Parameter estimation with ``colorful'' noise} \label{sec:noise}
  
Consider an experiment in which samples of an observable $y_i$ are
recorded at a sequence of times $\{t_i: i=1,\ldots,N \}$.  In the
context of a transit light curve, $y_i$ is the relative brightness of
the host star.  We assume that the times $t_i$ are known with
negligible error.  We further assume that in the absence of noise, the
samples $y_i$ would be given by a deterministic function,
\begin{eqnarray}
 	y(t_i) &=& f(t_i; p_1, \ldots, p_K) = f(t_i;\vec{p}),~~~{\rm (no~noise)}
\end{eqnarray}
where $\vec{p} = \{p_1,\ldots, p_K\}$ is a set of $K$ parameters that
specify the function $f$. For an idealized transit light curve, those
parameters may be the fractional loss of light $\delta$, the total
duration $T$, and ingress or egress duration $\tau$, and the
midtransit time $t_c$, in the notation of Carter et al.~(2008). More
realistic functions have been given by Mandel \& Agol (2002) and
Gim\'enez (2007).

We further suppose that a stochastic noise process $\epsilon(t)$ has
been added to the data, giving
\begin{eqnarray}
  	y(t_i) = f(t_i; \vec{p})+\epsilon(t_i).~~~{\rm (with~noise)} \label{eq:modelgen}
\end{eqnarray}
As a stochastic function, $\vec{ \epsilon} = \{\epsilon(t_1),\ldots
\epsilon(t_N)\}$ is characterized by its joint distribution function
${\cal D}(\vec{\epsilon}; \vec{q})$, which in turn depends on some
parameters $\vec{q}$ and possibly also the times of observation. The
goal of parameter estimation is to use the data $y(t_i)$ to calculate
credible intervals for the parameters $\vec{p}$, often reported as
best estimates $\hat{p}_k$ and error bars $\hat{\sigma}_{p_k}$ with
some quantified degree of confidence. The estimate of $\vec{p}$ and
the associated errors depend crucially on how one models the noise and
how well one can estimate the relevant noise parameters $\vec{q}$.

In some cases one expects and observes the noise to be
uncorrelated. For example, the dominant noise source may be shot
noise, in which case the noise process is an uncorrelated Poisson
process that in the limit of large numbers of counts is
well-approximated by an uncorrelated Gaussian process,
\begin{eqnarray}
{\cal D}(\vec{\epsilon}; \vec{q}) =
{\cal N}({\epsilon};{\sigma^2}) = \prod_{i=1}^{N} \frac{1}{\sqrt{2 \pi \sigma^2}}
   \exp \left(-\frac{\epsilon_i^2}{2 \sigma^2}\right), \label{eq:gaussian}
\end{eqnarray}
in which case there is only one error parameter, $\sigma$, specifying
the width of the distribution.

If the noise is correlated then it is characterized by a joint
probability distribution that is generally a function of all the times
of observation. We assume that the function is a multivariate Gaussian
function, in which case the noise process is entirely characterized by
the covariance matrix
\begin{eqnarray}
{\bf \Sigma}(t_i,t_j) = \langle \epsilon(t_i) \epsilon(t_j) \rangle. \label{eq:covariance-matrix}
\end{eqnarray}
Here, the quantity $\langle \epsilon \rangle$ is the mean of the
stochastic function $\epsilon$ over an infinite number of independent
realizations. We further assume that the covariance depends only on
the difference in time between two samples, and not on the absolute
time of either sample. In this case, the noise source is said to be
stationary and is described entirely by its autocovariance $R(\tau)$
(Bracewell 1965):
\begin{eqnarray}
R(\tau) \equiv \langle \epsilon(t) \epsilon(t+\tau) \rangle. \label{eq:autocovariance}
\end{eqnarray}

The parameter estimation problem is often cast in terms of finding the
set of parameters $\hat{p}_k$ that maximize a likelihood function. For
the case of Gaussian {\it uncorrelated}\, noise the likelihood
function is
\begin{eqnarray}
 {\cal L} &=&
 \prod_{i=1}^N \frac{1}{ \sqrt{2 \pi \hat{\sigma}^2} }
 \exp\left( -\frac{r_i^2}{2 \hat{\sigma}^2} \right), \label{eq:lwhite}
\end{eqnarray}
where $r_i$ is the {\it residual} defined as $y_i - f(t_i;\vec{p})$,
and $\hat{\sigma}$ is an estimate of the single noise parameter
$\sigma$.  Maximizing the likelihood ${\cal L}$ is equivalent to
minimizing the $\chi^2$ statistic
\begin{eqnarray}
  \chi^2 = \sum_i^N \left(  \frac{r_i}{\hat{\sigma}} \right)^2. \label{eq:chi2white}
\end{eqnarray}
In transit photometry, the estimator $\hat{\sigma}$ of the noise
parameter $\sigma$ is usually {\it not} taken to be the calculated
noise based on expected sources such as shot noise. This is because
the actual amplitude of the noise is often greater than the calculated
value due to noise sources that are unknown or at least
ill-quantified. Instead, $\hat{\sigma}$ is often taken to be the
standard deviation of the data obtained when the transit was not
occurring, or the value for which $\chi^2=N_{\rm dof}$ for the
best-fitting (minimum-$\chi^2$) model. These estimates work well when
the noise process is Gaussian, stationary, and uncorrelated. For the
case of correlated noise, Eqn.~(\ref{eq:chi2white}) is replaced by
(Gould 2003)
\begin{eqnarray}
   \chi^2 =  \sum_{i=1}^N \sum_{j=1}^N r_i (\hat{\Sigma}^{-1})_{ij} r_j. \label{eq:like}
\end{eqnarray}
The case of uncorrelated noise corresponds to $\hat{\Sigma}_{ij} =
\hat{\sigma}^2 \delta_{ij}$.

It is at this point where various methods for modeling correlated
noise begin to diverge. One approach is to estimate $\hat{\Sigma}$
from the sample autocovariance $\hat{R}(\tau)$ of the time series,
just as $\hat{\sigma}$ can be estimated from the standard deviation of
the residuals in the case of uncorrelated noise. However, the
calculation of $\chi^2$ has a worst-case time-complexity of $O(N^2)$
and iterative parameter estimation techniques can be prohibitively
slow. One might ameliorate the problem by truncating the covariance
matrix at some maximum lag, i.e., by considering the truncated
$\chi^2$ statistic
\begin{eqnarray}
  \chi^2(L) =
   \sum_{i=1}^N \sum_{\stackrel{\scriptstyle l=-L}{1<i+l<N}}^L r_i
   (\hat{\Sigma}^{-1})_{i(i+l)} r_{i+l}, \label{eq:liketruncated}
\end{eqnarray}
but in the presence of long-range correlations one needs to retain
many lags to obtain accurate parameter estimates. (In
\S~\ref{sec:truecov}, we will give an example where 50--75 lags were
needed.) Alternatively, one may model the autocorrelation function and
therefore the covariance matrix using an autoregressive moving-average
(ARMA) model with enough terms to give a good fit to the data (see,
e.g., Koen \& Lombard~1993). Again, though, in the presence of
long-range correlations the model covariance matrix will be non-sparse
and computationally burdensome.

Pont et al.~(2006) presented a useful simplification in the context of
a transit search, when data are obtained on many different nights. In
such cases it is reasonable to approximate the covariance matrix as
block-diagonal, with different blocks corresponding to different
nights. Pont et al.~(2006) also gave a useful approximation for the
covariance structure within each block, based on the variance in
boxcar-averaged versions of the signal. Ultimately their procedure
results in an equation resembling Eqn.~(\ref{eq:chi2white}) for each
block, but where $\hat{\sigma}$ is the quadrature sum of $\sigma_w$
(the ``white noise'') and $\sigma_r$ (the ``red noise,'' estimated
from the boxcar-averaged variance). In this paper, all our examples
involve a single time series with stationary noise properties, and the
net effect of the Pont et al.~(2006) method is to enlarge the
parameter errors by a factor
\begin{equation}
\beta = \sqrt{ 1 + \left(\frac{\sigma_r}{\sigma_w}\right)^2 }
\end{equation}
relative to the case of purely white noise ($\sigma_r = 0$). We will
refer to this method as the ``time-averaging'' method.

Another approach is to use Eqn.~(\ref{eq:chi2white}) without any
modifications, but to perform the parameter optimization on a large
collection of simulated data sets that are intended to have the same
covariance structure as the actual data set. This is the basis of the
``residual permutation'' method that is also discussed further in
\S~\ref{sec:comparison}.  As mentioned above, this method is a variant
of a bootstrap analysis that takes into account time-correlated noise.
More details on both the time-averaging and residual-permutation
methods are given in \S~\ref{sec:comparison}.

Our approach in this paper was motivated by the desire to allow for
the possibility of long-range correlations, and yet to avoid the
slowness of any method based on Eqn.~(\ref{eq:liketruncated}) or other
time-domain methods. Rather than characterizing the noise in the time
domain, we characterize it by its Power Spectral Density (PSD) ${\cal
  S}(f)$ at frequency $f$, defined as the square of the Fourier
transform of $\epsilon(t)$, or equivalently, the Fourier transform of
the autocovariance $R(\tau)$.  We restrict our discussion to noise
sources with a PSD
\begin{eqnarray}
	{\cal S}(f) = \frac{\SPSD}{f^\gamma}
\end{eqnarray}
for some $\SPSD > 0$ and spectral index $\gamma$. For the special case
of uncorrelated noise, $\gamma = 0$ and ${\cal S}(f)$ is independent
of $f$. This type of noise has equal power density at all frequencies,
which is why it is called ``white noise,'' in an analogy with visible
light.  As $\gamma$ is increased, there is an increasing preponderance
of low-frequency power over high-frequency power, leading to
longer-range correlations in time.

\begin{figure*}[htbp] 
  \epsscale{0.8}
  \plotone{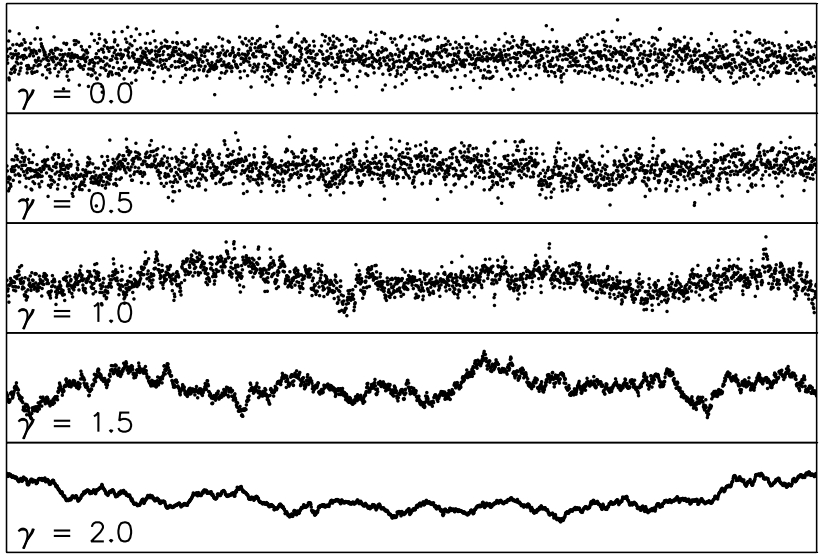}
  \caption{Examples of $1/f^\gamma$ noise. Uncorrelated (white) noise
    corresponds to $\gamma = 0$. ``Pink'' noise corresponds to $\gamma
    = 1$. ``Red'' noise or Brownian motion corresponds to $ \gamma =
    2$. These time series were generated using the wavelet-based
    method described in \S~\ref{sec:analysis}.}
  \label{fig:noises}
\end{figure*}

Noise with a power spectrum $1/f^\gamma$ is ubiquitous in nature and
in experimental science, including astrophysics (see, e.g., Press
1978). Some examples of $1/f^\gamma$ noise are shown in
Fig.~\ref{fig:noises} for a selection of spectral indices. In an
extension of the color analogy, $\gamma = 1$ noise is sometimes
referred to as ``pink noise'' and $\gamma = 2$ noise as ``red noise.''
The latter is also known as a Brownian process, although not because
of the color brown but instead because of the Scottish botanist Robert
Brown. However, as we have already noted, the term ``red noise'' is
often used to refer to any type of low-frequency correlated noise.
 
Here we do not attempt to explain how $1/f^\gamma$ noise arises in a
given situation. Instead we assume that the experimenter has done his
or her best to understand and to reduce all sources of noise as far as
possible, but despite these efforts there remains a component of
$1/f^\gamma$ noise. In transit photometry these correlations often
take the form of ``bumps,'' ``wiggles,'' and ``ramps'' in a light
curve and are often attributed to differential atmospheric extinction,
instrumental artifacts such as imperfect flat-fielding, and stellar
granulation or other astrophysical effects. The method presented in
this paper is essentially a model of the likelihood function that
retains the essential information in the covariance matrix without
being prohibitively expensive to compute and store. It is based on a
wavelet-based description, the subject of the next section.
  
\section{Wavelets and \lowercase{$1/f^\gamma$} noise} \label{sec:wavelet}
  
One may regard a time series with $N$ points as a vector in an
$N$-dimensional space that is spanned by $N$ orthonormal unit vectors,
one for each time index (the ``time basis''). The computational
difficulty with correlated noise is that the sample covariance matrix
$\hat{\Sigma}$ is not diagonal in the time basis, nor is it
necessarily close to being diagonal in realistic cases. This motivates
a search for some alternative basis spanning the data space for which
the covariance matrix is diagonal or nearly diagonal. For example, if
the noise took the form of additive quasiperiodic signals, it would be
logical to work in a Fourier basis instead of the time basis.

The mathematical result that underpins our analysis algorithm is that
in the presence of $1/f^\gamma$ noise, the covariance matrix is nearly
diagonal in a suitable {\it wavelet basis}.  Before giving the details
of the algorithm we will briefly review the wavelet transform. Our
discussion is drawn primarily from Wornell (1996), Teolis (1998),
Daubechies (1988), and Mallat (1999). Practical details and an sample
implementation of the wavelet transform are given by Press et
al.~(2007).

A wavelet is a function that is analogous to the sine and cosine
functions of the Fourier transform. Some properties that wavelets
share with sines and cosines are that they are localized in frequency
space, and they come in families that are related by translations and
dilations. Wavelets are {\em unlike} sine and cosine functions in that
wavelets are strongly localized in time. A wavelet basis is derived
from a single ``mother wavelet'' $\psi(t)$, which may have a variety
of functional forms and analytic properties. The individual basis
functions are formed through translations and dilations of $\psi(t)$.
The choice of mother wavelet depends on the specific application. We
restrict our focus to dyadic orthogonal wavelet bases with basis
functions
\begin{eqnarray}
  \psi_n^m(t) = \psi(2^{m}t-n) \label{eq:wavelet}
\end{eqnarray}
for all integers $m$ and $n$, and we further require $\psi(t)$ to have
one or more vanishing moments.\footnote{In particular it is required
  that the mother wavelet $\psi(t)$ has zero mean. This is a necessary
  and sufficient condition to ensure the invertibility of the wavelet
  transform.} In this case, the pair of equations analogous to the
Fourier series and its inversion is
\begin{eqnarray}
	\epsilon(t) & = & \sum_{m=-\infty}^{\infty} \sum_{n=-\infty}^{\infty} \epsilon_n^m  \psi_n^m(t) \label{eq:inverse} \\
	\epsilon_n^m &= & \int_{-\infty}^\infty \epsilon(t) \psi_n^m(t) dt \label{eq:wavelet-transform}
\end{eqnarray}
where $\epsilon_n^m$ is referred to as the wavelet coefficient of
$\epsilon(t)$ at resolution $m$ and translation $n$.

\subsection{The wavelet transform as a multiresolution analysis}

We will see shortly that some extra terms are required in
Eqn.~(\ref{eq:wavelet-transform}) for real signals with some minimum
and maximum resolution. To explain those terms it is useful to
describe the wavelet transform as a multiresolution analysis, in which
we consider successively higher-resolution approximations of a
signal. An approximation with a resolution of $2^m$ samples per unit
time is a member of a {\it resolution space} $V_m$. Following Wornell
(1996) we impose the following conditions:
\begin{enumerate}
   \item if $f(t) \in V_m$ then for some integer $n$, $f(t-2^{-m}n) \in V_m$
   \item if $f(t) \in V_m$ then $f(2 t) \in V_{m+1}$. 
\end{enumerate}
The first condition requires that $V_m$ contain all translations (at
the resolution scale) of any of its members, and the second condition
ensures that the sequence of resolutions is nested: $V_m$ is a subset
of the next finer resolution $V_{m+1}$. In this way, if $\epsilon_m(t)
\in V_m$ is an approximation to the signal $\epsilon(t)$, then the
next finer approxmation $\epsilon_{m+1}(t) \in V_{m+1}$ contains all
the information encoded in $\epsilon_m(t)$ plus some additional
{\it detail} $d_m(t)$ defined as
\begin{equation}
	d_m(t) \equiv \epsilon_{m+1}(t) - \epsilon_{m}(t).
\end{equation}
We may therefore build an approximation at resolution $M$ by starting
from some coarser resolution $k$ and adding successive detail
functions:
\begin{eqnarray}
	\epsilon_M(t) & = & \epsilon_k(t)+\sum_{m=k}^{M} d_m(t) \label{eq:dwaveletinverse}
\end{eqnarray}
The detail functions $d_m(t)$ belong to a function space $W_m(t)$, the
orthogonal complement of the resolution $V_m$.

With these conditions and definitions, the orthogonal basis functions
of $W_m$ are the wavelet functions $\psi_n^m(t)$, obtained by
translating and dilating some mother wavelet $\psi(t)$. The orthogonal
basis functions of $V_m$ are denoted $\phi_n^m(t)$, obtained by
translating and dilating a so-called ``father'' wavelet
$\phi(t)$. Thus, the mother wavelet spawns the basis of the detail
spaces, and the father wavelet spawns the basis of the resolution
spaces. They have complementary characteristics, with the mother
acting as a high-pass filter and the father acting as a low-pass
filter.\footnote{More precisely, the wavelet and scaling functions
  considered here are ``quadrature mirror filters'' (Mallat 1999).}

In Eqn.~(\ref{eq:dwaveletinverse}), the approximation $\epsilon_k(t)$
is a member of $V_k$, which is spanned by the functions $\phi_n^k(t)$,
and $d_m(t)$ is a member of $W_m$, which is spanned by the functions
$\psi_n^m(t)$. Thus we may rewrite Eqn.~(\ref{eq:dwaveletinverse}) as
\begin{eqnarray}
	\epsilon_M(t)  & = &
     \sum_{n=-\infty}^{\infty} \bar{\epsilon}_n^k \phi_n^k (t) + 
     \sum_{m=k}^{M} \sum_{n=-\infty}^{\infty} \epsilon_n^m \psi_n^m(t). \label{eq:approx}
\end{eqnarray}
The {\it wavelet coefficients} $\epsilon_n^m$ and the {\it scaling
  coefficients} $\bar{\epsilon}_n^m$ are given by
\begin{eqnarray}
	\epsilon_n^m &=& \int_{-\infty}^{\infty} \epsilon(t) \psi_n^m(t) dt \\
	\bar{\epsilon}_n^m &=& \int_{-\infty}^{\infty} \epsilon(t) \phi_n^m(t) dt
\end{eqnarray}
Eqn.~(\ref{eq:approx}) reduces to the wavelet-only
equation~(\ref{eq:inverse}) for the case of a continuously sampled
signal $\epsilon(t)$, when we have access to all resolutions $m$ from
$-\infty$ to $\infty$.\footnote{The signal must also be bounded in
  order for the approximation to the signal at infinitely coarse
  resolution to vanish, i.e., $\lim_{k\rightarrow-\infty}
  \epsilon_k(t) = 0$.}

\begin{figure*}[htbp] 
   \epsscale{1.0}
    \plottwo{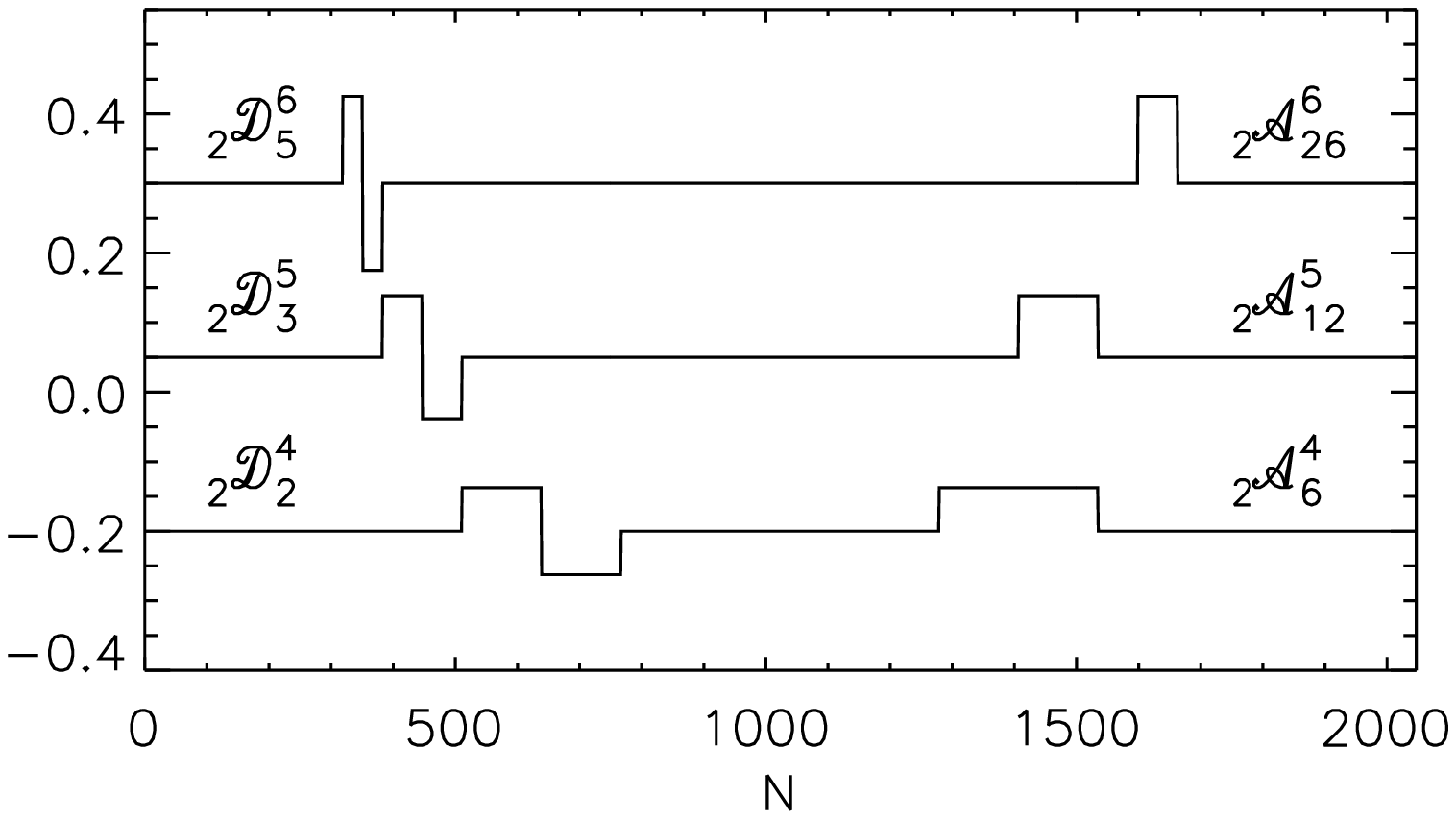}{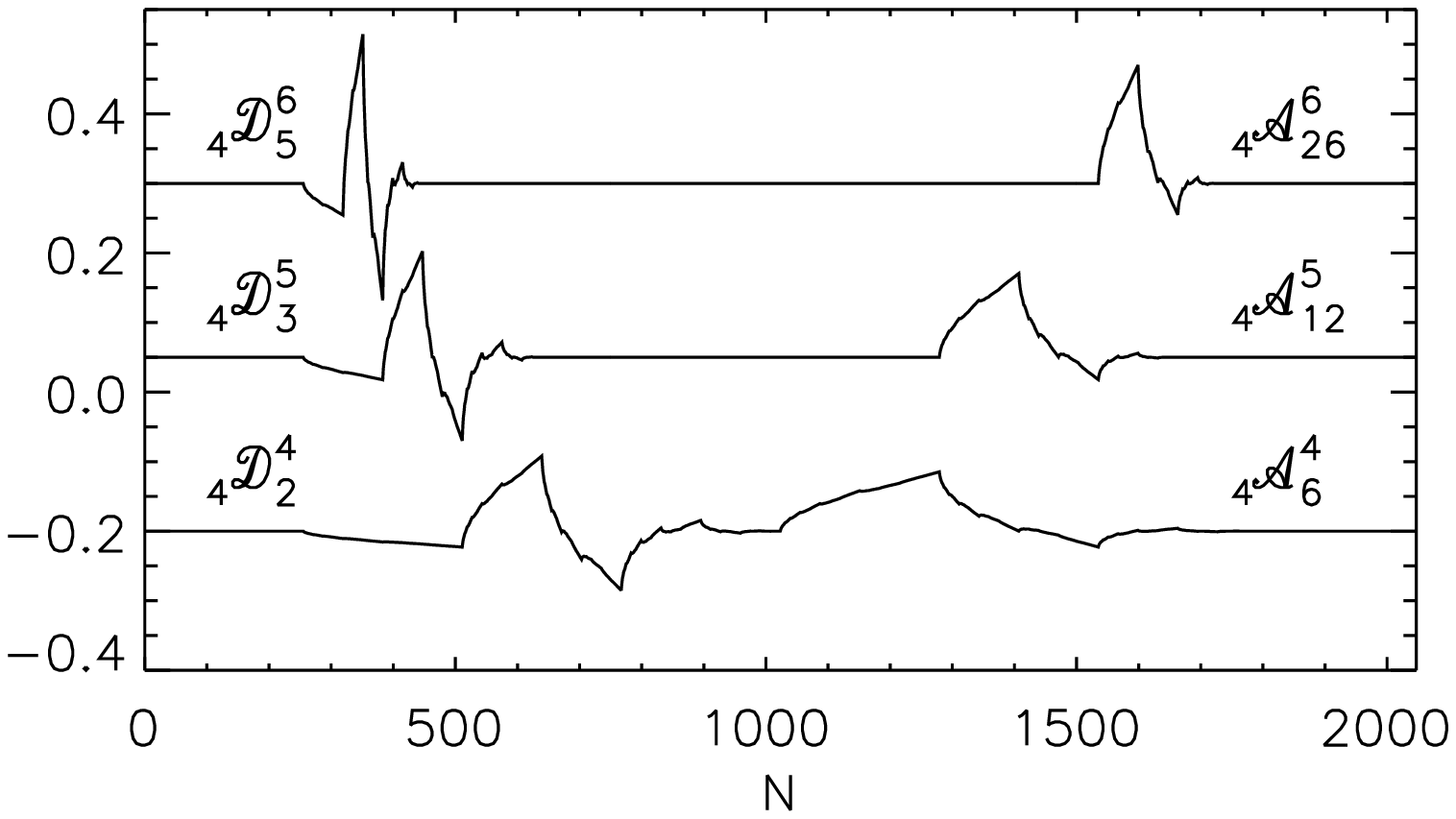}
    \caption{Examples of discrete wavelet and scaling functions, for
      $N=2048$.  {\it Left.}---Haar wavelets and the corresponding
      father wavelets, also known as 2nd-order Daubechies
      orthonormal wavelets or $_2\!{\cal D}_n^m$ and $_2\!{\cal
        A}_n^m$. {\it Right.}---4th-order Daubechies orthonormal
      wavelets, or $_4\!{\cal D}_n^m$ and $_4\!{\cal A}_n^m$.}
    \label{fig:d4}
\end{figure*}

\begin{figure*}[htbp] 
   \epsscale{1.0}
    \plottwo{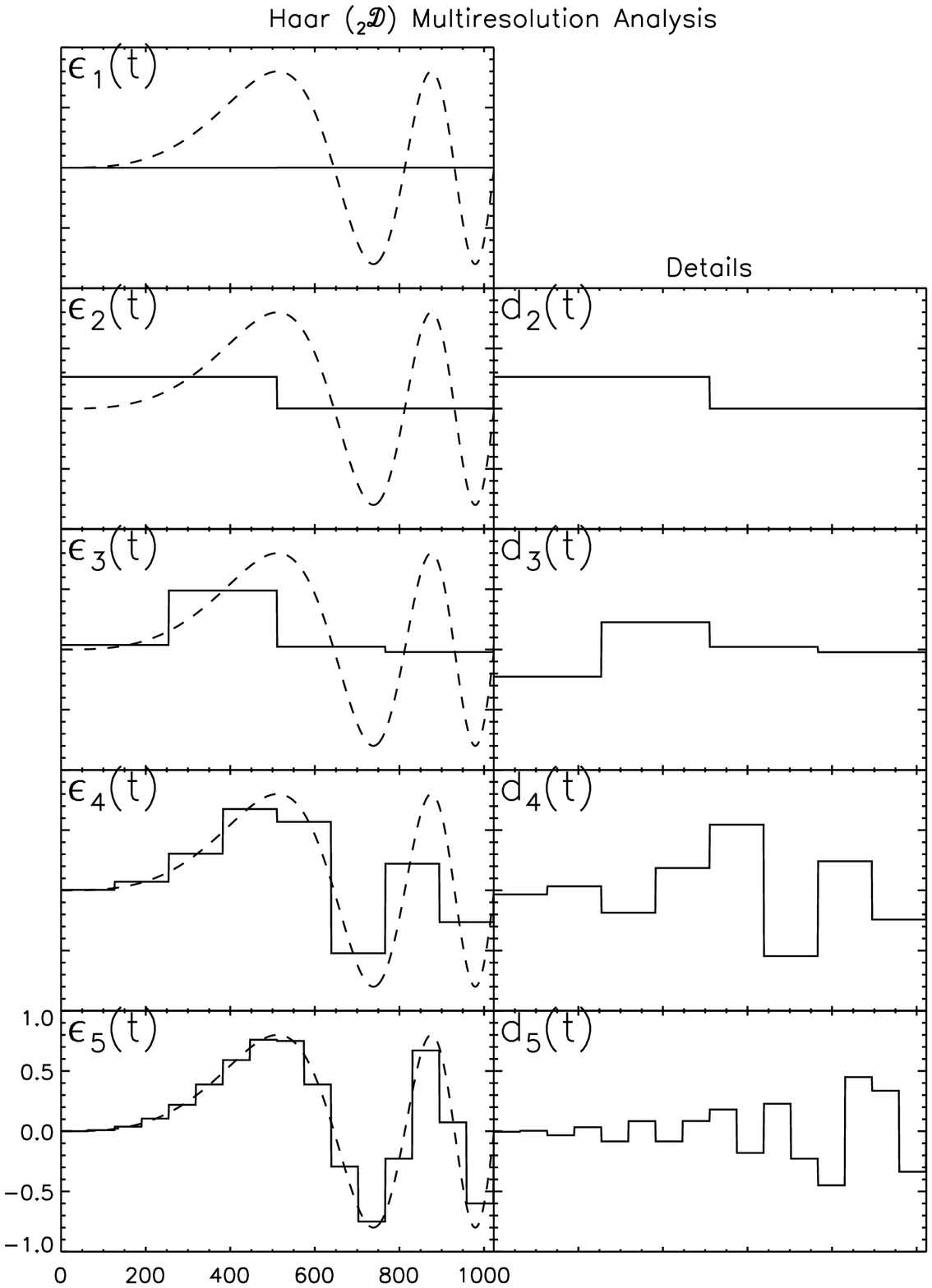}{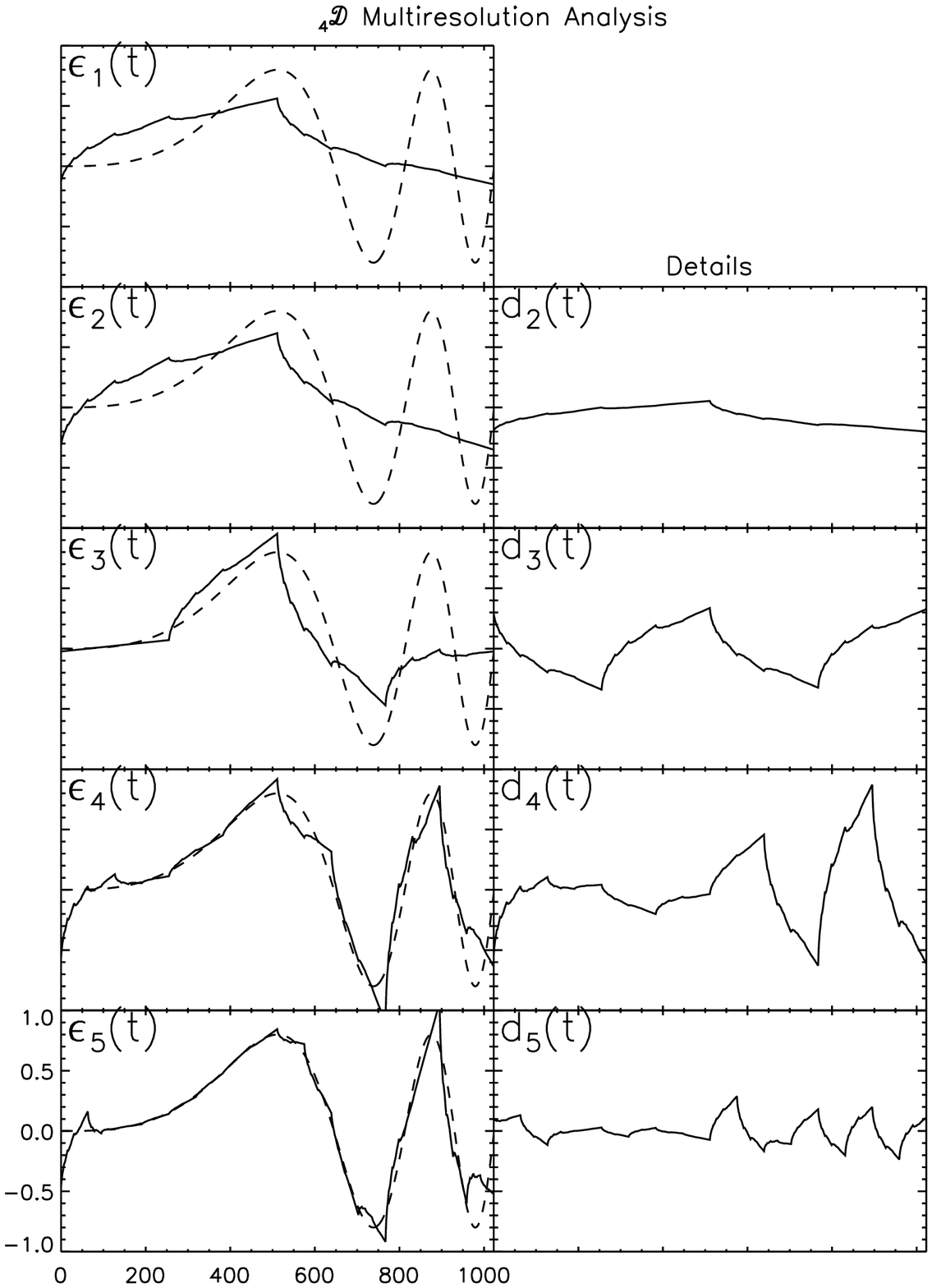}
    \caption{Illustration of a multiresolution analysis, for the
      function $\epsilon(t) = \sin[4 \pi (t/1024)^3]$ (dashed line).
      Plotted are the approximations $\epsilon_m(t)$ to the function
      at successive resolutions, along with the detail functions
      $d_m(t)$.  {\it Left.}---Using the Haar wavelet basis.  {\it
        Right.}---Using the 4th-order Daubechies wavelet basis.}
    \label{fig:mra}
\end{figure*}

There are many suitable choices for $\phi$ and $\psi$, differing in
the tradeoff that must be made between smoothness and localization.
The simplest choice is due to Haar (1910):
\begin{eqnarray}
	\phi(t) &=& \left\{ \begin{array}{ll}
			1\;\;\; & \mbox{if $0 < t \le 1$} \\
			0\;\;\; & \mbox{otherwise}
			\end{array}
		\right. . \\
	\psi(t) &=& \left\{ \begin{array}{ll}
			1\;\;\; & \mbox{if $-\frac{1}{2} < t \le 0$} \\
			-1\;\;\; & \mbox{if $0 < t \le \frac{1}{2}$} \\
			0\;\;\; & \mbox{otherwise}
			\end{array}
		\right.
\end{eqnarray}
The left panel of Fig.~\ref{fig:d4} shows several elements of the
approximation and detail bases for a Haar multiresolution analysis.
The left panels of Fig.~\ref{fig:mra} illustrate a Haar
multiresolution analysis for an arbitrarily chosen signal
$\epsilon(t)$, by plotting both the approximations $\epsilon_m(t)$ and
details $d_m(t)$ at several resolutions $m$. The Haar analysis is
shown for pedagogic purposes only. In practice we found it
advantageous to use the more complicated fourth-order Daubechies
wavelet basis, described in the next section, for which the elements
and the multiresolution analysis are illustrated in the right panels
of Fig.~\ref{fig:d4}-\ref{fig:mra}.

\subsection{The Discrete Wavelet Transform} 

Real signals are limited in resolution, leading to finite $M$ and $k$
in Eqn.~(\ref{eq:approx}). They are also limited in time, allowing
only a finite number of translations $N_m$ at a given resolution
$m$. Starting from Eqn.~(\ref{eq:approx}), we truncate the sum over
$n$ and reindex the resolution sum such that the coarsest resolution
is $k=1$, giving
\begin{eqnarray}
      \epsilon_M(t)  & = &
      \sum_{n=1}^{N_1} \bar{\epsilon}_n^1 \phi_n^1 (t) + 
      \sum_{m=2}^{M} \sum_{n=1}^{N_m} \epsilon_n^m \psi_n^m(t)
\end{eqnarray}
where we have taken $t=0$ to be the start of the signal. Since there
is no information on timescales smaller than $2^{-M}$, we need only
consider $\epsilon_M(t_i)$ at a finite set of times $t_i$:
\begin{eqnarray}
      \epsilon(t_i)  & = &
      \sum_{n=1}^{N_1} \bar{\epsilon}_n^1 \phi_n^1(t_i) + 
      \sum_{m=2}^{M} \sum_{n=1}^{N_m} \epsilon_n^m \psi_n^m(t_i) \label{eq:dwt}.
\end{eqnarray}
Eqn.~(\ref{eq:dwt}) is the inverse of the Discrete Wavelet Transform
(DWT). Unlike the continuous transform of Eqn.~(\ref{eq:inverse}), the
DWT must include the coarsest level approximation (the first term in
the preceding equation) in order to preserve all the information in
$\epsilon(t_i)$. For the Haar wavelet, the coarsest approximation is
the mean value. For data sets with $N = n_0 2^M$ uniformly spaced
samples in time, we will have access to a maximal scale $M$, as in
Eqn.~(\ref{eq:dwt}), with $N_m = n_0 2^{m-1}$.

A crucial point is the availability of the Fast Wavelet Transform
(FWT) to perform the DWT (Mallat 1989).  The FWT is a pyramidal
algorithm operating on data sets of size $N=n_0 2^M$ returning $n_0
(2^M-1)$ wavelet coefficients and $n_0$ scaling coefficients for some
$n_0 > 0$, $M > 0$.  The FWT is a computationally efficient algorithm
that is easily implemented (Press et al.\ 2007) and has $O(N)$
time-complexity (Teolis 1998).

Daubechies (1988) generalized the Haar wavelet into a larger family of
wavelets, categorized according to the number of vanishing moments of
the mother wavelet. The Haar wavelet has a single vanishing moment and
is the first member of the family. In this work we used the most
compact member (in time and frequency), $\psi = _4\!{\cal D}$ and
$\phi = _4\!{\cal A}$, which is well suited to the analysis of
$1/f^\gamma$ noise for $0 < \gamma < 4$ (Wornell 1996). We plot $_4
\!{\cal D}_n^m$ and $_4\!{\cal A}_n^m$ in the time-domain for several
$n$, $m$ in Fig.~\ref{fig:d4}, illustrating the rather unusual
functional form of $_4 \!{\cal D}$. The right panel of
Fig.~\ref{fig:mra} demonstrates a multiresolution analysis using
this basis. Press et al.\ (2007) provide code to implement the wavelet
transform in this basis.

\subsection{Wavelet transforms and $1/f^\gamma$ noise}

As alluded in \S~\ref{sec:wavelet}, the wavelet transform acts as a
nearly diagonalizing operator for the covariance matrix in the
presence of $1/f^\gamma$ noise. The wavelet coefficients
$\epsilon_n^m$ of such a noise process are zero-mean, nearly
uncorrelated random variables. Specifically, the covariance between
scales $m$, $m'$ and translations $n$, $n'$ is (Wornell 1996, p.~65)
\begin{eqnarray}
  \langle \epsilon_n^{m} \epsilon_{n'}^{m'} \rangle & \approx &
  \left( \SIGR 2^{-\gamma m} \right) \delta_{m,m'} \delta_{n,n'}. \label{eq:covnowhite}
\end{eqnarray}

The wavelet basis is also convenient for the case in which the noise
is modeled as the sum of an uncorrelated component and a correlated
component,
\begin{eqnarray}
	\epsilon(t) & =& \epsilon_0(t) + \epsilon_{\gamma}(t), \label{eq:nmodel}
\end{eqnarray}
where $\epsilon_0(t)$ is a Gaussian white noise process ($\gamma=0$)
with a single noise parameter $\sigma_w$, and $\epsilon_\gamma(t)$ has
${\cal S}(f) = \SPSD/f^\gamma$. In the context of transit photometry,
white noise might arise from photon-counting statistics (and in cases
where the detector is well-calibrated, $\sigma_w$ is a known
constant), while the $\gamma\neq 0$ term represents the ``rumble'' on
many time scales due to instrumental, atmospheric, or astrophysical
sources. For the noise process of Eqn.~(\ref{eq:nmodel}) the
covariance between wavelet coefficients is
\begin{eqnarray} 
  \langle \epsilon_n^{m} \epsilon_{n'}^{m'} \rangle & \approx &
  \left( \sigma_r^2 2^{-\gamma m} + \sigma_w^2 \right) \delta_{m,m'} \delta_{n,n'}. \label{eq:covwithwhite}
\end{eqnarray}
and the covariance betwen the scaling coefficients
$\bar{\epsilon}_n^m$ is
\begin{eqnarray}
  \langle \bar{\epsilon}_n^m \bar{\epsilon} _n^m\rangle & \approx &
  \SIGR 2^{-\gamma m} g(\gamma) + \sigma_w^2 \label{eq:varscale}
\end{eqnarray}
where $g(\gamma)$ is a constant of order unity; for the purposes of
this work $g(1) = (2 \ln 2)^{-1} \approx 0.72$ (Fadili \& Bullmore
2002). Eqns.~(\ref{eq:covwithwhite}) and (\ref{eq:varscale}) are the
key mathematical results that form the foundation of our
algorithm. For proofs and further details, see Wornell (1996).

It should be noted that the correlations between the wavelet and
scaling coefficients are small but not exactly zero. The decay rate of
the correlations with the resolution index depends on the choice of
wavelet basis and on the spectral index $\gamma$. By picking a wavelet
basis with a higher number of vanishing moments, we hasten the decay
of correlations. This is why we chose the Daubechies 4th-order basis
instead of the Haar basis. In the numerical experiments decribed in
\S~4, we found the covariances to be negligible for the purposes of
parameter estimation. In addition, the compactness of the Daubechies
4th-order basis reduces artifacts arising from the assumption of a
periodic signal that is implicit in the FWT.


\subsection{The whitening filter} \label{sec:whitening}
 
Given an observation of noise $\epsilon(t)$ that is modeled as in
Eqn.~(\ref{eq:nmodel}), we may estimate the $\gamma \ne 0$ component
by rescaling the wavelet and scaling coefficients and filtering out
the white component:
\begin{eqnarray}
  \epsilon_{\gamma}(t) &=& \sum_{n=1}^{N_1} \left( \frac{\sigma_r^2 2^{-\gamma} g(\gamma)}{\sigma_r^2 2^{-\gamma}g(\gamma)+\sigma_w^2}\right)\bar{\epsilon}_n^1 \phi_n^1(t)+ \\
	&& \sum_{m=2}^{M} \sum_{n=1}^{N_m} \left( \frac{\sigma_r^2 2^{-\gamma m}}{\sigma_r^2 2^{-\gamma m}+\sigma_w^2}\right) \epsilon_n^m\psi_n^m(t). \label{eq:filter}
\end{eqnarray}
We may then proceed to subtract the estimate of the correlated
component from the observed noise, $\epsilon_0(t) =
\epsilon(t)-\epsilon_\gamma(t)$ (Wornell 1996, p.~76). In this way the
FWT can be used to ``whiten'' the noise.
 
\subsection{The wavelet-based likelihood}
 
Armed with the preceding theory, we rewrite the likelihood function of
Eqn.~(\ref{eq:lwhite}) in the wavelet domain. First
we transform the residuals $r_i \equiv y_i-f(t_i; \vec{p})$, giving
\begin{eqnarray}
  r_n^m & = & y_n^m-f_n^m(\vec{p}) = \epsilon_{\gamma,n}^m+\epsilon_{0,n}^m \\
  \bar{r}_n^1 & = & \bar{y}_n^1-\bar{f}_n^1(\vec{p}) = \bar{\epsilon}_{\gamma,n}^1+\bar{\epsilon}_{0,n}^1 
\end{eqnarray}
where $y_n^m$ and $f_n^m(\vec{p})$ are the discrete wavelet
coefficients of the data and the model. Likewise, $\bar{y}_n^1$ and
$\bar{f}_n^1(\vec{p})$ are the $n_0$ scaling coefficients of the data
and the model.  Given the diagonal covariance matrix shown in
Eqns.~(\ref{eq:covwithwhite}) and (\ref{eq:varscale}), the likelihood
${\cal L}$ is a product of Gaussian functions at each scale $m$ and
translation $n$:
\begin{eqnarray}
	{\cal L} &=& \left\{ \prod_{m=2}^{M} \prod_{n=1}^{n_0 2^{m-1}} \frac{1}{\sqrt{2 \pi \sigma_W^2} }\exp\left[ -\frac{\left(r_n^m \right)^2}{2 \sigma_W^2}\right]  \right\}\nonumber \\
	&\times&\left\{ \prod_{n=1}^{n_0}  \frac{1}{\sqrt{2 \pi \sigma_S^2} }\exp\left[ -\frac{\left(\bar{r}_n^1\right)^2}{2 \sigma_S^2}\right] \right\}\label{eq:lwave}
\end{eqnarray}
where 
\begin{eqnarray}
	\sigma_W^2 & = & \SIGR 2^{-\gamma m}+\sigma_w^2 \\
	\sigma_S^2 & = & \SIGR 2^{-\gamma} g(\gamma)+\sigma_w^2
\end{eqnarray}
are the variances of the wavelet and scaling coefficients
respectively. For a data set with $N$ points, calculating the
likelihood function of Eqn.~(\ref{eq:lwave}) requires multiplying $N$
Gaussian functions. The additional step of computing the FWT of the
residuals prior to computing ${\cal L}$ adds $O(N)$ operations. Thus,
the entire calculation has a time-complexity $O(N)$.

For this calculation we must have estimators of the three noise
parameters $\gamma$, $\sigma_r$ and $\sigma_w$. These may be estimated
separately from the model parameters $\vec{p}$, or simultaneously with
the model parameters. For example, in transit photometry, the data
obtained outside of the transit may be used to estimate the noise
parameters, which are then used in Eqn.~(\ref{eq:lwave}) to estimate
the model parameters.  Or, in a single step we could maximize
Eqn.~(\ref{eq:lwave}) with respect to all of $\gamma$, $\sigma_r$,
$\sigma_w$ and $\vec{p}$. Fitting for both noise and transit
parameters simultaneously is potentially problematic, because some of
the correlated noise may be ``absorbed'' into the choices of the
transit parameters, i.e., the errors in the noise parameters and
transit parameters are themselves correlated. This may cause the noise
level and the parameter uncertainties to be
underestimated. Unfortunately, there are many instances when one does
not have enough out-of-transit data for the strict separation of
transit and noise parameters to be feasible.

In practice the optimization can be accomplished with an iterative
routine [such as AMOEBA, Powell's method, or a conjugate-gradient
method; see Press et al.~(2007)].  Confidence intervals can then be
defined by the contours of constant likelihood. Alternatively one can
use a Monte Carlo Markov Chain [MCMC; see, e.g., Gregory (2005)], in
which case the jump-transition likelihood would be given by
Eqn.~(\ref{eq:lwave}). The advantages of the MCMC method have led to
its adoption by many investigators (see, e.g., Holman et al.~2006,
Burke et al.~2007, Collier Cameron et al.~2007). For that method,
computational speed is often a limiting factor, as a typical MCMC
analysis involves several million calculations of the likelihood
function.

\subsection{Some practical considerations}

Some aspects of real data do not fit perfectly into the requirements
of the DWT. The time sampling of the data should be approximately
uniform, so that the resolution scales of the multiresolution analysis
accurately reflect physical timescales. This is usually the case for
time-series photometric data. Gaps in a time series can be fixed by
applying the DWT to each uninterrupted data segment, or by filling in
the missing elements of the residual series with zeros.

The FWT expects the number of data points to be an integral multiple
of some integral power of two. When this is not the case, the time
series may be truncated to the nearest such boundary; or it may be
extended using a periodic boundary condition, mirror reflection, or
zero-padding. In the numerical experiments described below, we found
that zero-padding has negligible effects on the calculation of
likelihood ratios and parameter estimation.

The FWT generally assumes a periodic boundary condition for simplicity
of computation. A side effect of this simplication is that information
at the beginning and end of a time series are artificially associated
in the wavelet transform. This is one reason why we chose the
4th-order Daubechies-class wavelet basis, which is well localized in
time, and does not significantly couple the beginning and the end of
the time series except on the coarsest scales.

\section{Numerical experiments with transit light curves} \label{sec:analysis}
  
We performed many numerical experiments to illustrate and test the
wavelet method. These experiments involved estimating the parameters
of simulated transit light curves. We also compared the wavelet
analysis to a ``white'' analysis, by which we mean a method that
assumes the errors to be uncorrelated, and to two other analysis
methods drawn from the literature. Because we used simulated transit
light curves with known noise and transit parameters, the ``truth''
was known precisely, allowing both the absolute and relative merits of
the methods to be evaluated.

\subsection{Estimating the midtransit time: Known noise parameters} \label{sec:known}

In this section we consider the case in which the noise parameters
$\gamma$, $\sigma_r$, and $\sigma_w$ are known with negligible
error. We have in mind a situation in which a long series of
out-of-transit data are available, with stationary noise properties.

We generated transit light curves with known transit parameters
$\vec{p}$, contaminated by an additive combination of a white and a
correlated ($1/f^\gamma$) noise source. Then we used an MCMC method to
estimate the transit parameters and their 68.3\% confidence limits.
(The technique for generating noise and the MCMC method are described
in detail below.) For each realization of a simulated light curve, we
estimated transit parameters using the likelihood defined either by
Eqn.~(\ref{eq:lwhite}) for the white analysis, or
Eqn.~(\ref{eq:lwave}) for the wavelet analysis.

For a given parameter $p_k$, the estimator $\hat{p}_k$ was taken to be
the median of the values in the Markov chain and $\hat{\sigma}_{p_k}$
was taken to be the standard deviation of those values. To assess the
results, we considered the ``number-of-sigma'' statistic
\begin{eqnarray}
  {\cal N} & \equiv \left(\hat{p}_k-p_k\right)/\hat{\sigma}_{p_k}. \label{eq:Np}
\end{eqnarray}
In words, ${\cal N}$ is the number of standard deviations separating
the parameter estimate $\hat{p}_k$ from the true value $p_k$. If the
error in $p_k$ is Gaussian, then a perfect analysis method should
yield results for ${\cal N}$ with an expectation value of $0$ and
variance of $1$.  If we find that the variance of ${\cal N}$ is
greater than one, then we have underestimated the error in $\hat{p}_k$
and we may attribute too much significance to the result. On the other
hand, if the variance of ${\cal N}$ is smaller than one, then we have
overestimated $\sigma_{p_k}$ and we may miss a significant
discovery. If we find that the mean of ${\cal N}$ is nonzero then the
method is biased.

For now, we consider only the single parameter $t_c$, the time of
midtransit. The $t_c$ parameter is convenient for this analysis as it
is nearly decoupled from the other transit parameters (Carter et al.\
2008). Furthermore, as mentioned in the introduction, the measurement
of the midtransit time cannot be improved by observing other transit
events, and variations in the transit interval are possible signs of
additional gravitating bodies in a planetary system.

The noise was synthesized as follows. First, we generated a sequence
of $N = 1024$ independent random variables obeying the variance
conditions from Eqns.~(\ref{eq:covwithwhite}) and (\ref{eq:varscale})
for $1023$ wavelet coefficients over $9$ scales and a single scaling
coefficient at the coarsest resolution scale.  We then performed the
inverse FWT of this sequence to generate our noise signal.  In this
way, we could select exact values for $\gamma$, $\sigma_r$, and
$\sigma_w$. We also needed to find the single parameter $\sigma$ for
the white-noise analysis; it is not simply related to the parameters
$\gamma$, $\sigma_r$, and $\sigma_w$. In practice, we found $\sigma$
by calculating the median sample variance among $10^4$ unique
realizations of a noise source with fixed parameters $\gamma$,
$\sigma_r$, and $\sigma_w$.

\begin{figure*}[htbp] 
   \epsscale{1.0}
    \plotone{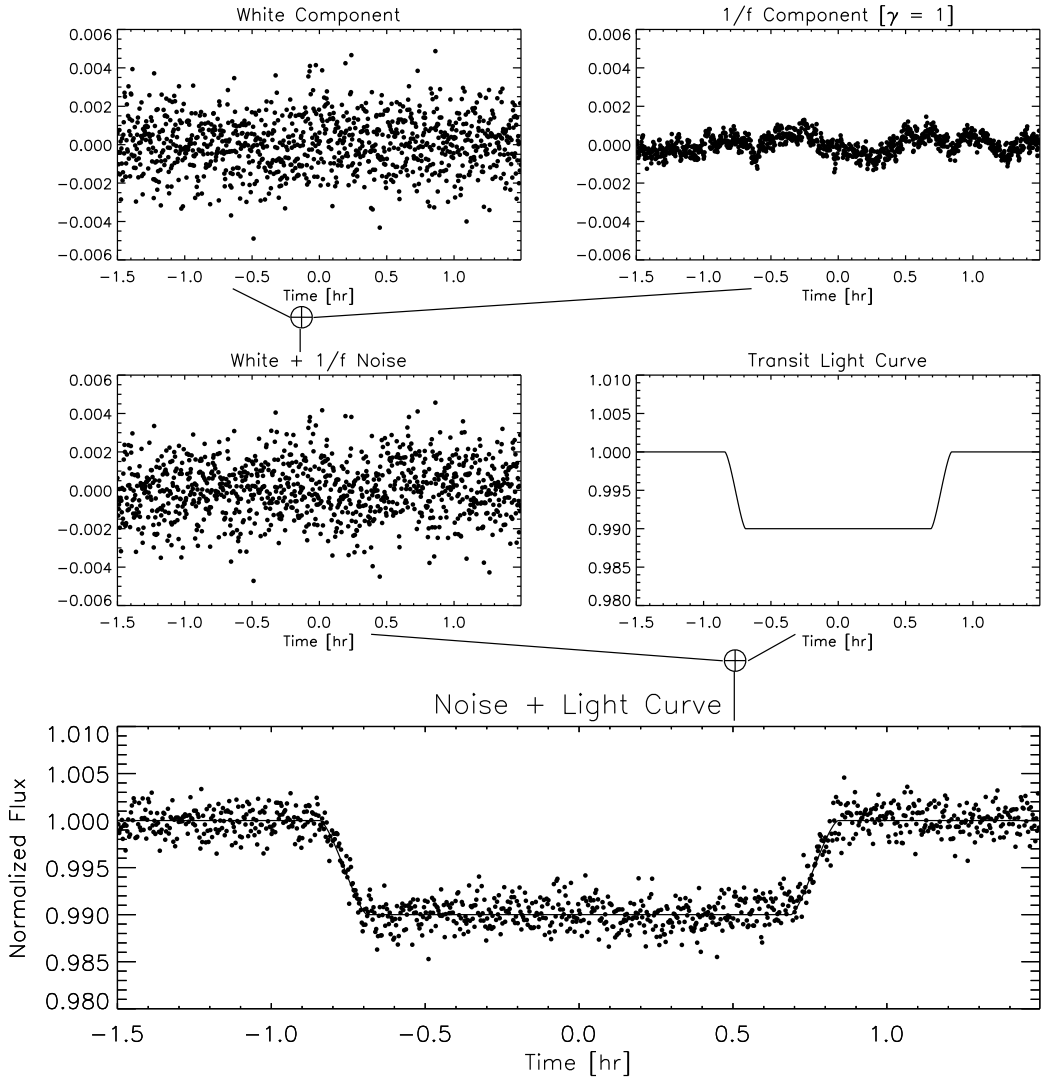}
    \caption{Constructing a simulated transit light curve with
      correlated noise.  The total noise is the sum of uncorrelated
      Gaussian noise with standard deviation $\sigma_w$ (upper left
      panel) and correlated noise with a power spectral density $S(f)
      \propto 1/f$ and an rms equal to $\sigma_w/3$ (upper right
      panel). The total noise (middle left panel) is added to an
      idealized transit model (middle right panel) to produce the
      simulated data (bottom panel).}
    \label{fig:sample_noise}
\end{figure*}

For the transit model, we used the analytic formulas of Mandel \& Agol
(2002), with a planet-to-star ratio of $R_p/R_\star = 0.15$, a
normalized orbital distance of $a/R_\star = 10$, and an orbital
inclination of $i=90\arcdeg$, as appropriate for a gas giant planet in
a close-in orbit around a K star. These correspond to a fractional
loss of light $\delta=0.0225$, duration $T=1.68$~hr, and partial
duration $\tau=0.152$~hr. We did not include the effect of limb
darkening, as it would increase the computation time and has little
influence on the determination of $t_c$ (Carter et al.~2009). Each
simulated light curve spanned 3~hr centered on the midtransit time,
with a time sampling of 11~s, giving 1024 uniformly spaced samples. A
noise-free light curve is shown in Fig.~\ref{fig:sample_noise}.

\begin{figure*}[htbp] 
   \epsscale{1.0}
    \plotone{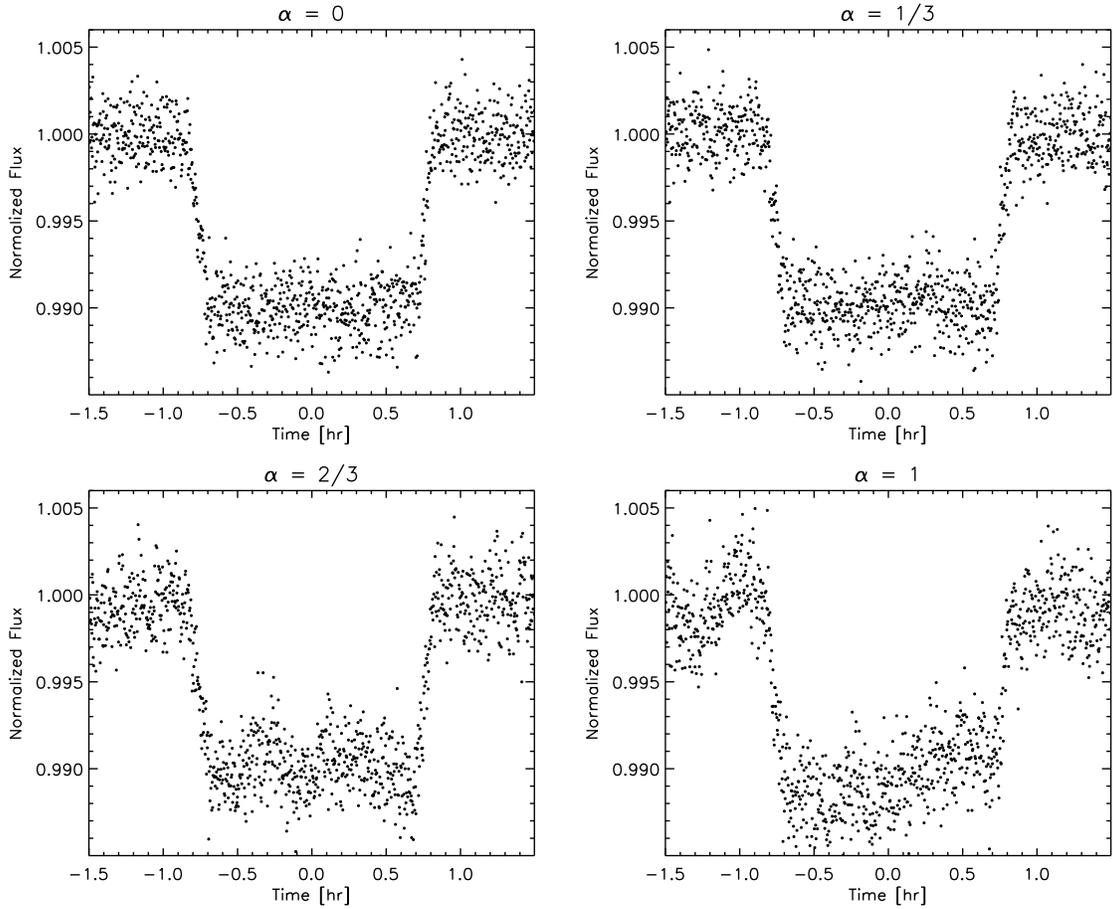}
    \caption{Examples of simulated transit light curves with different
      ratios $\alpha = {\rm rms}_r / {\rm rms}_w$ between the rms
      values of the correlated noise component and white noise
      component.}
    \label{fig:sample_noise2}
\end{figure*}

For the noise model, we chose $\sigma_w = 1.35 \times 10^{-3}$ and
$\gamma = 1$, and tried different choices for $\sigma_r$.  We denote
by $\alpha$ the ratio of the rms values of the correlated noise
component and the white noise component.\footnote{We note that
  although $\sigma_w$ is the rms of the white noise component,
  $\sigma_r$ is generally not the rms of the correlated component. The
  notation is unfortunate, but follows that of Wornell~(1996).} The
example in Fig.~\ref{fig:sample_noise} has $\alpha=1/3$. As $\alpha$
is increased from zero, the correlated component becomes more
important, as is evident in the simulated data plotted in
Fig.~\ref{fig:sample_noise2}. Our choice of $\sigma_w$ corresponds to
a precision of $5.8\times 10^{-4}$ per minute-equivalent sample, and
was inspired by the recent work by Johnson et al.~(2009) and Winn et
al.~(2009), which achieved precisions of $5.4\times 10^{-4}$ and
$4.0\times 10^{-4}$ per minute-equivalent sample, respectively. Based
on our survey of the literature and our experience with the Transit
Light Curve project (Holman et al.~2006, Winn et al.~2007), we submit
that all of the examples shown in Fig.~\ref{fig:sample_noise2} are
``realistic'' in the sense that the bumps, wiggles, and ramps resemble
features in actual light curves, depending on the instrument,
observing site, weather conditions, and target star.

For a given choice of $\alpha$, we made 10,000 realizations of the
simulated transit light curve with $1/f$ noise. We then constructed
two Monte Carlo Markov Chains for $t_c$ starting at the true value of
$t_c = 0$. One chain was for the white analysis, with a
jump-transition likelihood given by Eqn.~(\ref{eq:lwhite}). The other
chain was for the wavelet analysis, using Eqn.~(\ref{eq:lwave})
instead. Both chains used the Metropolis-Hastings jump condition, and
employed perturbation sizes such that $\approx$40\% of jumps were
accepted. Initial numerical experiments showed that the
autocorrelation of a given Markov chain for $t_c$ is sharply peaked at
zero lag, with the autocorrelation dropping below $0.2$ at
lag-one. This ensured good convergence with chain lengths of $500$
(Tegmark et al.~2004). Chain histograms were also inspected visually
to verify that the distribution was smooth. We recorded the median
$\hat{t}_c$ and standard deviation $\hat{\sigma}_{t_c}$ for each chain
and constructed the statistic ${\cal N}$ for each separate analysis
(white or wavelet). Finally, we found the median and standard
deviation of ${\cal N}$ over all 10,000 noise realizations.

\begin{figure*}[htbp] 
   \epsscale{0.8}
    \plotone{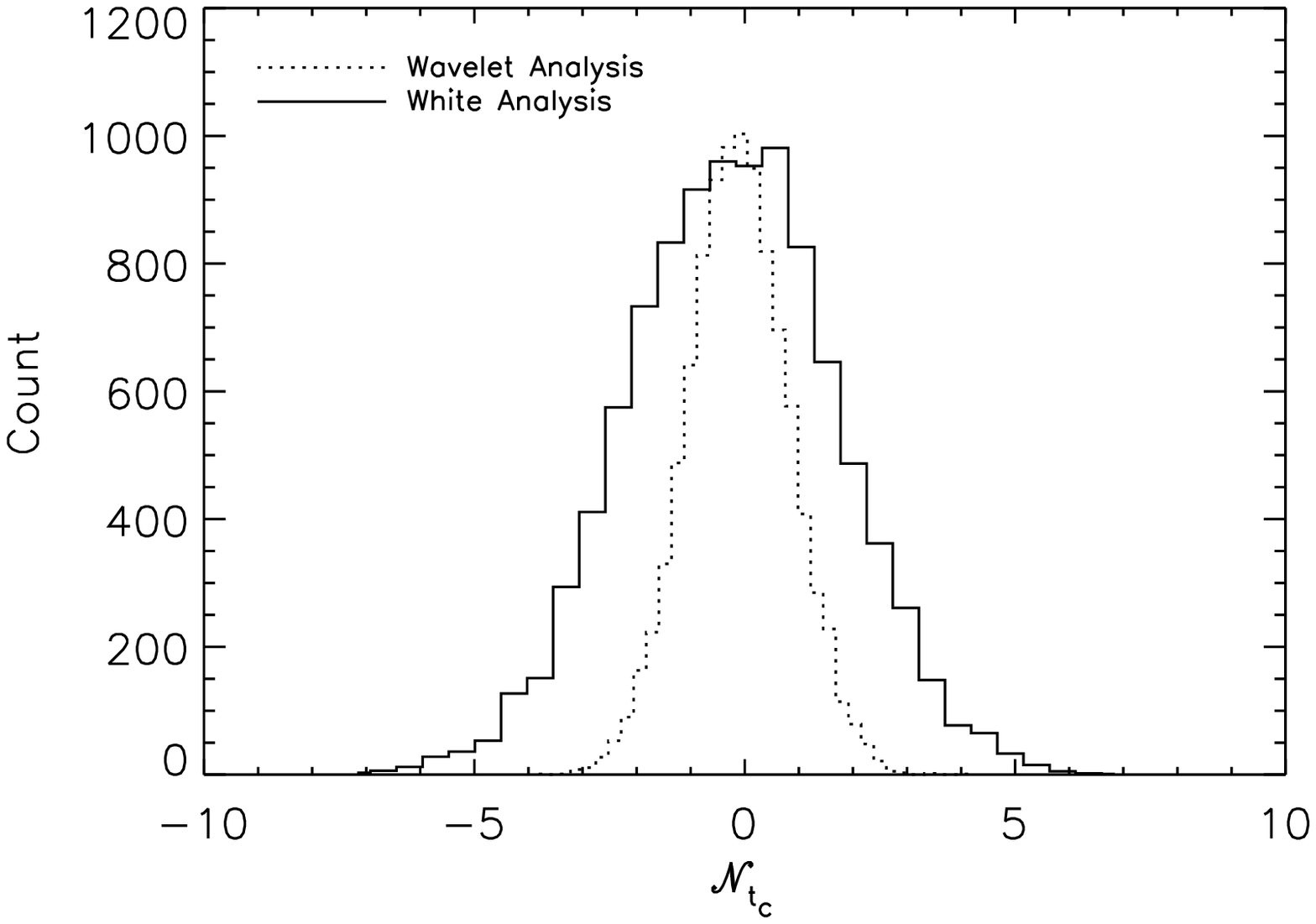}
    \caption{Histograms of the number-of-sigma statistic ${\cal N}$
      for the midtransit time $t_c$. Each distribution shows the
      probability of estimating a value for $t_c$ that differs by
      ${\cal N}$$\sigma$ from the true value. The simulated data were
      created by adding an idealized transit model to a noise source
      that is the sum of uncorrelated noise and $1/f$ noise with equal
      variances ($\alpha=1$; see the text). \label{fig:ntcdist}}
\end{figure*}

Fig.~\ref{fig:ntcdist} shows the resulting distributions of ${\cal
  N}$, for the particular case $\alpha = 1/3$. Table \ref{tab:known}
gives a collection of results for the choices $\alpha = 0$, $1/3$,
$2/3$, and $1$. The mean of ${\cal N}$ is zero for both the white and
wavelet analyses: neither method is biased. This is expected, because
all noise sources were described by zero-mean Gaussian
distributions. However, the widths of the distributions of ${\cal N}$
show that the white analysis underestimates the error in $t_c$.  For a
transit light curve constructed with equal parts white and $1/f$ noise
($\alpha = 1$), the white analysis gave an estimate of $t_c$ that
differs from the true value by more than $1~\sigma$ nearly $80\%$ of
the time. The factor by which the white analysis underestimates the
error in $t_c$ appears to increase linearly with $\alpha$. In
contrast, for all values of $\alpha$, the wavelet analysis maintains a
unit variance in ${\cal N}$, as desired.

The success of the wavelet method is partially attributed to the
larger (and more appropriate) error intervals that it returns for
$\hat{t}_c$. It is also partly attributable to an improvement in the
accuracy of $\hat{t}_c$ itself: the wavelet method tends to produce
$\hat{t}_c$ values that are closer to the true $t_c$.  This is shown
in the final column in Table~(\ref{tab:known}), where we report the
percentage of cases in which the analysis method (white or wavelet)
produces an estimate of $t_c$ that is closer to the truth.  For
$\alpha=1$ the wavelet analysis gives more accurate results $66\%$ of
the time.

\begin{deluxetable}{lcccccc}
\tablecolumns{7}
\tablewidth{0pt}
\tablecaption{Estimates of mid-transit time, $t_c$, from data with known noise properties}
\tablehead{\colhead{Method} & \colhead{$\alpha$} &
		\colhead{$\langle \hat{\sigma}_{t_c} \rangle$ [sec]} &
		\colhead{$\langle {\cal N} \rangle$} & \colhead{$\sigma_{{\cal N}}$} & \colhead{{\em prob}$({\cal N} > 1)$} & \colhead{{\em prob}$($best$)$\tablenotemark{a}} }

\startdata
\nl
White             &	$0$		&	$4.1$	&	$+0.004$	&	$0.95$	&	$29\%$	&	$50\%$	\nl
 		  &	$1/3$	&	$4.3$	&	$-0.005$	&	$1.93$	&	$61\%$	&	$39\%$	\nl
 		  &	$2/3$	&	$5.0$	&	$+0.005$	&	$3.04$	&	$75\%$	&	$35\%$	\nl		  
		  &	$1$		&	$5.9$	&	$-0.036$	&	$3.82$	&	$79\%$	&	$34\%$	\nl \nl \tableline \nl
Wavelet      &	$0$		&	$4.0$	&	$+0.005$	&	$0.95$	&	$29\%$	&	$50\%$	\nl
 		  &	$1/3$	&	$7.2$	&	$-0.004$	&	$0.93$	&	$28\%$	&	$61\%$	\nl
 		  &	$2/3$	&	$11.5$	&	$-0.004$	&	$0.94$	&	$28\%$	&	$65\%$	\nl
		  &	$1$		&	$16.0$	&	$-0.001$	&	$0.95$	&	$29\%$	&	$66\%$	\nl
\enddata
\tablenotetext{a}{The probability that the analysis method (white or
  wavelet) returns an estimate of $t_c$ that is closer to the true
  value than the other method.\label{tab:known} }
\end{deluxetable}

\subsection{Estimating the midtransit time: Unknown noise parameters} \label{sec:unknown}

In this section we consider the case in which the noise parameters are
not known in advance. Instead the noise parameters must be estimated
based on the data. We did this by including the noise parameters as
adjustable parameters in the Markov chains. In principle this could be
done for all three noise parameters $\gamma$, $\sigma_r$, and
$\sigma_w$, but for most of the experiments presented here we
restricted the problem to the case $\gamma=1$. This may be a
reasonable simplification, given the preponderance of natural noise
sources with $\gamma = 1$ (Press 1978). Some experiments involving
noise with $\gamma \ne 1$ are described at the end of this section.

We also synthesized the noise with a non-wavelet technique, to avoid
``stacking the deck'' in favor of the wavelet method. We generated the
noise in the frequency domain, as follows. We specified the amplitudes
of the Fourier coefficients using the assumed functional form of the
power spectral density [${\cal S}(f) \propto 1/f$], and drew the
phases from a uniform distribution between $-\pi$ and $\pi$. The
correlated noise in the time domain was found by performing an inverse
Fast Fourier Transform. We rescaled the noise such that the rms was
$\alpha$ times the specified $\sigma_w$. The normally-distributed
white noise was then added to the correlated noise to create the total
noise. This in turn was added to the idealized transit model.

For each choice of $\alpha$, we made 10,000 simulated transit light
curves and analyzed them with the MCMC method described previously.
For the white analysis, the mid-transit time $t_c$ and the single
noise parameter $\sigma$ were estimated using the likelihood defined
via Eqn.~(\ref{eq:lwhite}). For the wavelet analysis we estimated
$t_c$ and the two noise parameters $\sigma_r$ and $\sigma_w$ using the
likelihood defined in Eqn.~(\ref{eq:lwave}).

Table~\ref{tab:unknown} gives the resulting statistics from this
experiment, in the same form as were given in Table~\ref{tab:known}
for the case of known noise parameters. (This table also includes some
results from \S~\ref{sec:comparison}, which examines two other methods
for coping with correlated noise.) Again we find that the wavelet
method produces a distribution of ${\cal N}$ with unit variance,
regardless of $\alpha$; and again, we find that the white analysis
underestimates the error in $t_c$. In this case the degree of error
underestimation is less severe, a consequence of the additional
freedom in the noise model to estimate $\sigma$ from the data. The
wavelet method also gives more accurate estimates of $t_c$ than the
white method, although the contrast between the two methods is smaller
than it was with for the case of known noise parameters.

Our numerical results must be understood to be illustrative, and not
universal. They are specific to our choices for the noise parameters
and transit parameters. Via further numerical experiments, we found
that the width of ${\cal N}$ in the white analysis is independent of
$\sigma_w$, but it does depend on the time sampling. In particular,
the width grows larger as the time sampling becomes finer (see
Table~\ref{tab:samp}). This can be understood as a consequence of the
long-range correlations. The white analysis assumes that the increased
number of data points will lead to enhanced precision, whereas in
reality, the correlations negate the benefit of finer time sampling.

\begin{deluxetable}{ccc}
\tablecolumns{2}
\tablewidth{0pt}
\tablecaption{Effect of time sampling on the white analysis}
\tablehead{\colhead{$N$\tablenotemark{a}} & \colhead{Cadence [sec]} & \colhead{$\sigma_{{\cal N}}$}}
\startdata
\nl
256 & 42.2 & 1.72 \nl
512 &21.1 & 2.04 \nl
1024 &10.5 & 2.69 \nl
2048& 5.27 & 3.49 \nl
4096& 2.63 &  4.39	\nl
\enddata
\tablenotetext{a}{The number of samples in a 3~hr interval.}
\label{tab:samp}
\end{deluxetable}

Table~\ref{tab:unkgamma} gives the results of additional experiments
with $\gamma \ne 1$. In those cases we created simulated noise with
$\gamma \ne 1$ but in the course of the analysis we assumed
$\gamma=1$.  The correlated noise fraction was set to $\alpha = 1/2$
for these tests. The results show that even when $\gamma$ is falsely
assumed to be unity, the wavelet analysis still produces better
estimates of $t_c$ and more reliable error bars than the white
analysis.

\begin{deluxetable*}{lcccccc}
\tablecolumns{7}
\tablewidth{0pt}
\tablecaption{Estimates of $t_c$ from data with unknown noise properties}
\tablehead{\colhead{Method} & \colhead{$\alpha$} &
		\colhead{$\langle \hat{\sigma}_{t_c} \rangle$ [sec]} &
		\colhead{$\langle {\cal N} \rangle$} & \colhead{$\sigma_{{\cal N}}$} & \colhead{{\em prob}$({\cal N} > 1)$} & \colhead{{\em prob}$($better$)$\tablenotemark{a}} }

\startdata
\nl
White          &	$0$		&	$4.0$	&	$-0.011$	&	$0.97$	&	$31\%$	&	$-$	\nl
 		  &	$1/3$	&	$4.2$	&	$+0.010$	&	$1.70$	&	$57\%$	&	$-$	\nl
 		  &	$2/3$	&	$4.9$	&	$+0.012$	&	$2.69$	&	$73\%$	&	$-$	\nl		  
		  &	$1$		&	$5.8$	&	$+0.023$	&	$3.28$	&	$78\%$	&	$-$	\nl \nl \tableline \nl
Wavelet      &	$0$		&	$4.5$	&	$-0.009$	&	$0.90$	&	$26\%$	&	$50\%$	\nl
 		  &	$1/3$	&	$6.9$	&	$-0.003$	&	$1.03$	&	$33\%$	&	$56\%$	\nl
 		  &	$2/3$	&	$11.2$	&	$-0.005$	&	$1.07$	&	$35\%$	&	$57\%$	\nl
		  &	$1$		&	$15.7$	&	$-0.007$	&	$1.09$	&	$36\%$	&	$57\%$	\nl  \nl \tableline \nl
Time-averaging      &	$0$		&	$4.4$	&	$-0.006$	& $0.88$      &      $26\%$     &       $50\%$      \nl 
 		  	&	$1/3$	&	$6.8$	&	$+0.009$	&	$1.15$	&	$36\%$	&	$50\%$	\nl
 		 	 &	$2/3$	&	$11.6$	&	$-0.012$	&	$1.24$	&	$40\%$	&	$50\%$	\nl
		  	&	$1$		&	$17.6$	&	$+0.007$	&	$1.21$	&	$38\%$	&	$50\%$	\nl  \nl \tableline \nl
Residual-permutation     &	$0$		&	$3.5$	&	$-0.012$ &  $1.16$      &      $37\%$     &       $50\%$      \nl  
 					  &	$1/3$	&	$6.6$	&	$+0.013$	&	$1.24$	&	$37\%$	&	$50\%$	\nl
 					  &	$2/3$	&	$11.8$	&	$-0.014$	&	$1.28$	&	$38\%$	&	$49\%$	\nl
					  &	$1$		&	$17.3$	&	$+0.008$	&	$1.30$	&	$38\%$	&	$48\%$	\nl
\enddata
\tablenotetext{a}{The probability that the analysis method returns an estimate of $t_c$ that is closer to the true
  value than the white analysis.}
\label{tab:unknown}
\end{deluxetable*}

 \begin{deluxetable}{lcccccc}
\tablecolumns{7}
\tablewidth{0pt}
\tablecaption{Estimates of $t_c$ from data with unknown noise properties}
\tablehead{\colhead{Method} & \colhead{$\gamma$\tablenotemark{a}} &
		\colhead{$\langle \hat{\sigma}_{t_c} \rangle$ [sec]} &
		\colhead{$\langle {\cal N} \rangle$} & \colhead{$\sigma_{{\cal N}}$} & \colhead{{\em prob}$({\cal N} > 1)$} & \colhead{{\em prob}$($best$)$\tablenotemark{b}} }

\startdata
\nl
White          &	$0.5$	&	$4.5$	&	$-0.025$	&	$1.34$	&	$47\%$	&	$50\%$	\nl
 		  &	$1.5$	&	$4.6$	&	$+0.020$	&	$3.10$	&	$77\%$	&	$32\%$	\nl \nl \tableline \nl
Wavelet      &	$0.5$	&	$6.7$	&	$-0.021$	&	$0.97$	&	$30\%$	&	$50\%$	\nl
 		  &	$1.5$	&	$6.9$	&	$+0.002$	&	$1.17$	&	$39\%$	&	$68\%$	\nl
\enddata
\tablenotetext{a}{The spectral exponent of the Power Spectral Density, $S(f) \propto 1/f^\gamma$.}
\tablenotetext{b}{The probability that the analysis method (white or
  wavelet) returns an estimate of $t_c$ that is closer to the true
  value than the other method.}
\label{tab:unkgamma}
\end{deluxetable}

\subsection{Runtime analysis of the time-domain method} \label{sec:truecov}\nopagebreak

Having established the superiority of the wavelet method over the
white method, we wish to show that the wavelet method is also
preferable to the more straightforward approach of computing the
likelihood function in the time domain with a non-diagonal covariance
matrix. The likelihood in this case is given by Eqn.~(\ref{eq:like}).
 
The time-domain calculation and the use of the covariance matrix
raised two questions. First, how well can we estimate the
autocovariance $R(\tau)$ from a single time series? Second, how much
content of the resulting covariance matrix needs to be retained in the
likelihood calculation for reliable parameter estimation? The answer
to the first question depends on whether we wish to utilize the sample
autocorrelation as the estimator of $R(\tau)$ or instead use a
parametric model (such as an ARMA model) for the autocorrelation.  In
either case, our ability to estimate the autocorrelation improves with
number of data samples contributing to its calculation. The second
question is important because retaining the full covariance matrix
would cause the computation time to scale as $O(N^2)$ and in many
cases the analysis would be prohibitively slow. The second question
may be reframed as: what is the minimum number of lags $L$ that needs
to be considered in computing the truncated $\chi^2$ of
Eqn.~(\ref{eq:liketruncated}), in order to give unit variance in the
number-of-sigma statistic for each model parameter? The
time-complexity of the truncated likelihood calculation is $O(NL)$. If
$L\la 5$ then the time-domain method and the wavelet method may have
comparable computational time-complexity, while for larger $L$ the
wavelet method would offer significant advantage.

\begin{figure}[h] 
   \epsscale{1.0}
    \plotone{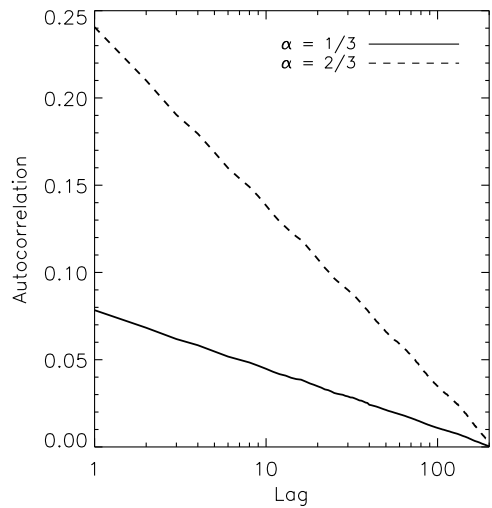}
    \caption{Autocorrelation functions of correlated noise. The noise
      was computed as the sum of white noise with $\sigma_w = 0.00135$
      and $1/f$ noise with an rms equal to $\alpha\sigma_w$, for
      $\alpha=1/3$ or $2/3$.}
    \label{fig:auto}
\end{figure}

We addressed these questions by repeating the experiments of the
previous sections using a likelihood function based on the truncated
$\chi^2$ statistic. We assumed that the parameters of the noise model
were known, as in \S~\ref{sec:known}. The noise was synthesized in the
wavelet domain, with $\gamma=1$, $\sigma_w=0.00135$, and $\alpha$ set
equal to $1/3$ or $2/3$. The parameters of the transit model and the
time series were the same as in \S~\ref{sec:known}.  We calculated the
``exact'' autocovariance function $R(l)$ at integer lag $l$ for a
given $\alpha$ by averaging sample autocovariances over 50,000 noise
realizations.  Fig.~\ref{fig:auto} plots the autocorrelation
[$R(l)/R(0)$] as a function of lag for $\alpha = 1/3$, $2/3$.  We
constructed the stationary covariance $\Sigma_{ij} = R(|i-j|)$ and
computed its inverse $(\Sigma^{-1})_{ij}$ for use in
Eqn.~(\ref{eq:liketruncated}).

\begin{figure}[h] 
   \epsscale{1.0}
    \plotone{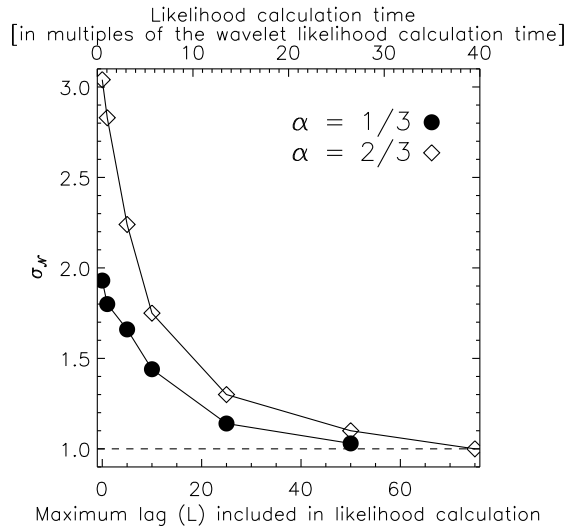}
    \caption{Accuracy of the truncated time-domain likelihood in
      estimating midtransit times. Plotted is the variance in the
      number-of-sigma statistic $\sigma_{{\cal N}}$ for the midtransit
      time $t_c$, as a function of the maximum lag in the truncated
      series. The estimates of $t_c$ were found using the truncated
      likelihood given in Eqn.~(\ref{eq:liketruncated}).  }
    \label{fig:trunc}
\end{figure}

Then we used the MCMC method to find estimates and errors for the time
of midtransit, and calculated the number-of-sigma statistic ${\cal N}$
as defined in Eqn.~(\ref{eq:Np}).  In particular, for each simulated
transit light curve, we created a Markov chain of $1,000$ links for
$t_c$, using $\chi^2(L)$ in the jump-transition likelihood. We
estimated $t_c$ and $\sigma_{t_c}$, and calculated ${\cal N}$. We did
this for $5,000$ realizations and determined $\sigma_{{\cal N}}$, the
variance in ${\cal N}$, across this sample.  We repeated this process
for different choices of the maximum lag $L$.  Fig.~(\ref{fig:trunc})
shows the dependence of $\sigma_{{\cal N}}$ upon the maximum lag $L$.

The time-domain method works fine, in the sense that when enough
non-diagonal elements in the covariance matrix are retained, the
parameter estimation is successful. We find that $\sigma_{{\cal N}}$
approaches unity as $L^{-\beta}$ with $\beta = 0.15$, $0.25$ for
$\alpha = 1/3$, $2/3$, respectively. However, to match the reliability
of the wavelet method, a large number of lags must be retained. To
reach $\sigma_{{\cal N}}$ = 1.05, we need $L \approx 50$ for
$\alpha = 1/3$ or $L \approx 75$ for $\alpha = 2/3$. In our
implementation, the calculation based on the truncated covariance
matrix [Eqn.~(\ref{eq:liketruncated})] took $30$--$40$ times longer
than the calculation based on the wavelet likelihood
[Eqn.~(\ref{eq:lwave})].

This order-of-magnitude penalty in runtime is bad enough, but the real
situation may be even worse, because one usually has access to a
single noisy estimate of the autocovariance matrix. Or, if one is
using an ARMA model, the estimated parameters of the model might be
subject to considerable uncertainty as compared to the ``exact''
autocovariance employed in our numerical experiments. If it is desired
to determine the noise parameters simultaneously with the other model
parameters, then there is a further penalty associated with inverting
the covariance matrix at each step of the calculation for use in
Eqn.~(\ref{eq:liketruncated}), although it may be possible to
circumvent that particular problem by modeling the inverse-covariance
matrix directly.

\subsection{Comparison with other methods} \label{sec:comparison}

In this section we compare the results of the wavelet method to two
methods for coping with correlated noise that are drawn from the
recent literature on transit photometry. The first of these two
methods is the ``time averaging'' method that was propounded by Pont
et al.~(2006) and used in various forms by Bakos et al.~(2006), Gillon
et al.~(2006), Winn et al.~(2007, 2008, 2009), Gibson et al.~(2008),
and others. In one implementation, the basic idea is to calculate the
sample variance of unbinned residuals, $\hat{\sigma}_{1}^2$, and also
the sample variance of the time-averaged residuals,
$\hat{\sigma}_{n}^2$, where every $n$ points have been averaged
(creating $m$ time bins). In the absence of correlated noise, we
expect
\begin{eqnarray}
  \hat{\sigma}_{n}^2 &=&\frac{ \hat{\sigma}_{1}^2}{n} \left(\frac{m}{m-1}\right).
\end{eqnarray}
In the presence of correlated noise, $\hat{\sigma}_{n}^2$ differs from
this expectation by a factor $\hat{\beta}_n^2$. The estimator
$\hat{\beta}$ is then found by averaging $\hat{\beta}_n$ over a range
$\Delta n$ corresponding to time scales that are judged to be most
important.  In the case of transit photometry, the duration of ingress
or egress is the most relevant time scale (corresponding to averaging
time scales on the order of tens of minutes, in our example light
curve). A white analysis is then performed, using the noise parameter
$\sigma = \beta\sigma_1$ instead of $\sigma_1$. This causes the
parameter errors $\hat{\sigma}_{p_k}$ to increase by $\beta$ but does
not change the parameter estimates $\hat{p}_k$
themselves.\footnote{Alternatively one may assign an error to each
  data point equal to the quadrature sum of the measurement error and
  an extra term $\sigma_r$ (Pont et al.~2006). For cases in which the
  errors in the data points are all equal or nearly equal, these
  methods are equivalent.  When the errors are not all the same, it is
  more appropriate to use the quadrature-sum approach of Pont et
  al.~(2006).  In this paper all our examples involve homogeneous
  errors.}

A second method is the ``residual permutation'' method that has been
used by Jenkins et al.~(2002), Moutou et al.~(2004),
Southworth~(2008), Bean et al.~(2008), Winn et al.~(2008), and
others. This method is a variant of a bootstrap analysis, in which the
posterior probability distribution for the parameters is based on the
collection of results of minimizing $\chi^2$ (assuming white noise)
for a large number of synthetic data sets. In the traditional
bootstrap analysis the synthetic data sets are produced by scrambling
the residuals and adding them to a model light curve, or by drawing
data points at random (with replacement) to make a simulated data set
with the same number of points as the actual data set. In the residual
permutation method, the synthetic data sets are built by performing a
cyclic permutation of the time indices of the residuals, and then
adding them to the model light curve. In this way, the synthetic data
sets have the same bumps, wiggles, and ramps as the actual data, but
they are translated in time. The parameter errors are given by the
widths of the distributions in the parameters that are estimated from
all the different realizations of the synthetic data, and they are
usually larger than the parameter errors returned by a purely white
analysis.

As before, we limited the scope of the comparison to the estimation of
$t_c$ and its uncertainty. We created 5,000 realizations of a noise
source with $\gamma=1$ and a given value of $\alpha$ (either $0$,
$1/3$, $2/3$, or $1$). We used each of the two approximate methods
(time-averaging and residual-permutation) to calculate $\hat{\beta}$
and its uncertainty based on each of the 5,000 noise
realizations. Then we found the median and standard deviation of
$\hat{\beta}/\beta$ over all 5,000 realizations.
Table~(\ref{tab:unknown}) presents the results of this experiment.

Both methods, time-averaging and residual-permutation, gave more
reliable uncertainties than the white method. However they both
underestimated the true uncertainties by approximately 15-30\%.
Furthermore, neither method provided more accurate estimates of $t_c$
than did the white method. For the time-averaging method as we have
implemented it, this result is not surprising, for that method differs
from the white method only in the inflation of the error bars by some
factor $\beta$. The parameter values that maximize the likelihood
function were unchanged.

\subsection{Alternative noise models} \label{sec:arma}

We have shown the wavelet method to work well in the presence of
$1/f^\gamma$ noise. Although this family of noise processes
encompasses a wide range of possibilities, the universe of possible
correlated noise processes is much larger. In this section we test the
wavelet method using simulated data that has correlated noise of a
completely different character. In particular, we focus on a process
with exclusively short-term correlations, described by one of the
aforementioned autoregressive moving-average (ARMA) class of
parametric noise models. In this way we test our method on a noise
process that is complementary to the longer-range correlations present
in $1/f^\gamma$ noise, and we also make contact between our method and
the large body of statistical literature on ARMA models.

For $1/f^\gamma$ noise we have shown that time-domain parameter
estimation techniques are slow. However, if the noise has exclusively
short-range correlations, the autocorrelation function will decay with
lag more rapidly than a power law, and the truncated-$\chi^2$
likelihood [Eqn.~(\ref{eq:liketruncated})] may become computationally
efficient. ARMA models provide a convenient analytic framework for
parameterizing such processes. For a detailed review of ARMA models
and their use in statistical inference, see Box \& Jenkins
(1976). Applications of ARMA models to astrophysical problems have
been described by in Koen \& Lombard (1993), Konig \& Timmer (1997)
and Timmer et al.\ (2000).

\begin{figure}[htbp] 
   \epsscale{1.0}
    \plotone{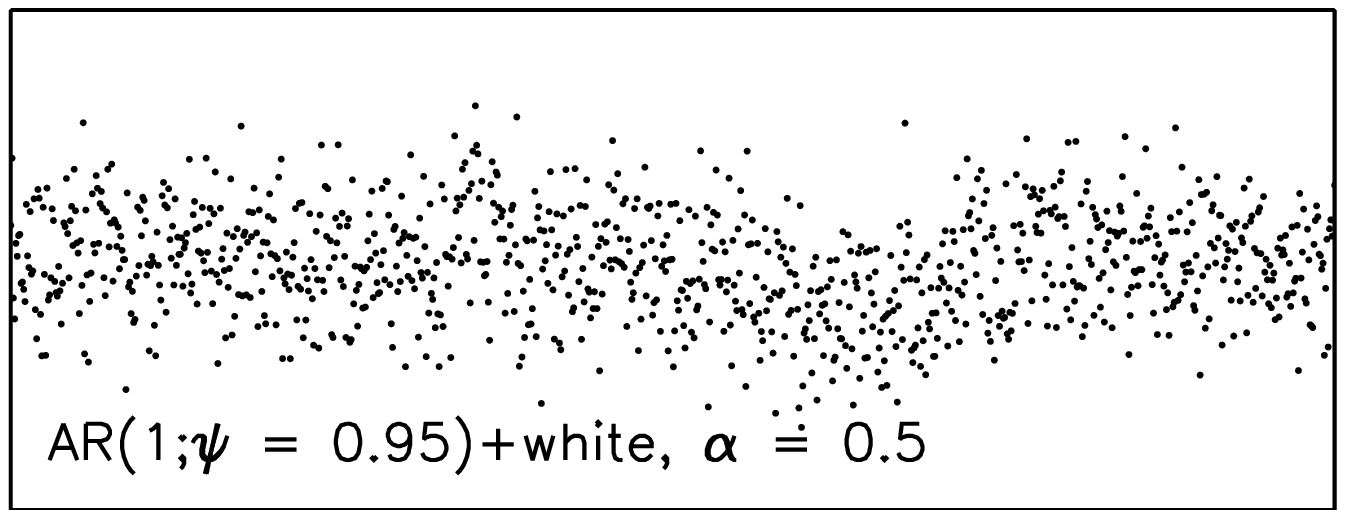}
    \plotone{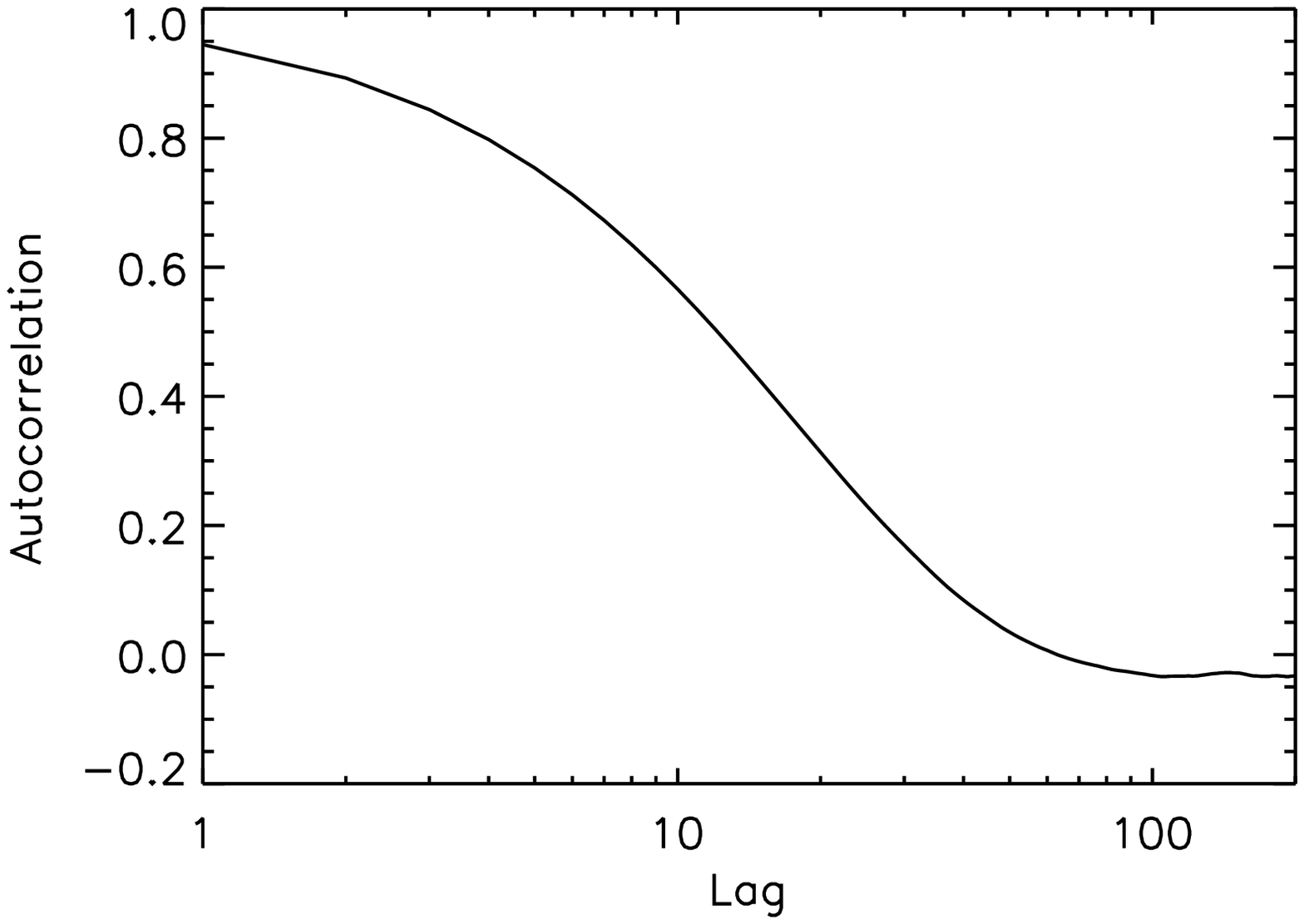}
    \caption{An example of an autoregressive noise process with
      complementary characteristics to a $1/f^\gamma$ process. The top
      panel shows the sum of an AR(1) process with $\psi=0.95$ and
      white noise.  The correlated and uncorrelated components have
      equal variances ($\alpha=0.5$). }
    \label{fig:ar}
\end{figure}

To see how the wavelet method performs on data with short-range
correlations we constructed synthetic transit data in which the noise
is described by a single-parameter autoregressive [AR(1;~$\psi$)]
model. An AR(1;~$\psi$) process $\epsilon(t_i)$ is defined by the
recursive relation
\begin{eqnarray}
	\epsilon(t_i) = \eta(t_i)+\psi \epsilon(t_{i-1})
\end{eqnarray}
where $\eta(t_i)$ is an uncorrelated Gaussian process with width
parameter $\sigma$ and $\psi$ is the sole autoregressive
parameter. The autocorrelation $\gamma(l)$ for an AR(1;~$\psi$)
process is
\begin{eqnarray}
	\gamma(l) = \frac{\sigma^2}{1-\psi^2} \psi^l.
\end{eqnarray}
An AR(1;$\psi$) process is stationary so long as $0 < \psi < 1$ (Box
\& Jenkins 1976). The decay length of the autocorrelation function
grows as $\psi$ is increased from zero to one. Figure~(\ref{fig:ar})
plots the autocorrelation function of a process that is an additive
combination of an AR(1;~$\psi=0.95$) process and a white noise
process. The noise in our synthetic transit light curves was the sum
of this AR(1;~$\psi = 0.95$) process, and white noise, with $\alpha =
1/2$ (see Fig~\ref{fig:ar}). With these choices, the white method
underestimates the error in $t_c$, while at the same time the
synthetic data look realistic.

We proceeded with the MCMC method as described previously to estimate
the time of mid-transit. All four methods assessed in the previous
section were included in this analysis, for
comparison. Table~\ref{tab:ar} gives the results. The wavelet method
produces more reliable error estimates than the white method.
However, the wavelet method no longer stands out as superior to the
time-averaging method or the residual-permutation method; all three of
these methods give similar results. This illustrates the broader point
that using any of these methods is much better than ignoring the noise
correlations. The results also show that although the wavelet method
is specifically tuned to deal with $1/f^\gamma$ noise, it is still
useful in the presence of noise with shorter-range correlations.

It is beyond the scope of this paper to test the applicability of the
wavelet method on more general ARMA processes. Instead we suggest the
following approach when confronted with real data [see also Beran
(1994)]. Calculate the sample autocorrelation, and power spectral
density, based on the out-of-transit data or the residuals to an
optimized transit model. For stationary processes these two indicators
are related as described in \S~\ref{sec:noise}. Short-memory,
ARMA-like processes can be identified by large autocorrelations at
small lags or by finite power spectral density at zero
frequency. Long-memory processes ($1/f^\gamma$) can be identified by
possibly small but non-vanishing autocorrelation at longer
lags. Processes with short-range correlations could be analyzed with
an ARMA model of the covariance matrix [see Box \& Jenkins (1976)], or
the truncated-lag covariance matrix, although a wavelet-based analysis
may be sufficient as well. Long-memory processes are best analyzed
with the wavelet method as described in this paper.

It should also be noted that extensions of ARMA models have been
developed to mimic long-memory, $1/f^\gamma$ processes. In particular,
fractional autoregressive integrated moving-average models (ARFIMA)
describe ``nearly'' $1/f^\gamma$ stationary processes, according to
the criterion described by Beran (1994). As is the case with ARMA
models, ARFIMA models enjoy analytic forms for the likelihood in the
time-domain. Alas, as noted by Wornell (1996) and Beran (1994), the
straightforward calculation of this likelihood is computationally
expensive and potentially unstable. For $1/f^\gamma$ processes, the
wavelet method is probably a better choice than any time-domain method
for calculating the likelihood.

\begin{deluxetable*}{lcccccc}
\tablecolumns{6}
\tablewidth{0pt}
\tablecaption{Estimates of $t_c$ from data with autoregressive correlated noise}
\tablehead{\colhead{Method}  &
		\colhead{$\langle \hat{\sigma}_{t_c} \rangle$ [sec]} &
		\colhead{$\langle {\cal N} \rangle$} & \colhead{$\sigma_{{\cal N}}$} & \colhead{{\em prob}$({\cal N} > 1)$} & \colhead{{\em prob}$($better$)$\tablenotemark{a}} }

\startdata
White          	&	$4.5$	&	$-0.010$	&	$2.50$	&	$70\%$	&	$-$	\nl
Wavelet     		&	$8.7$	&	$-0.016$	&	$1.33$	&	$44\%$	&	$51\%$	\nl
Time-averaging		&	$9.9$	&	$-0.010$	& $1.25$      &      $40\%$     &       $49\%$      \nl 
Residual-permutation     &	$10.2$	&	$-0.010$ &  $1.23$      &      $38\%$     &       $51\%$      \nl  

\enddata
\tablenotetext{a}{The probability that the analysis method returns an estimate of $t_c$ that is closer to the true
  value than the white analysis.}
\label{tab:ar}
\end{deluxetable*}

\subsection{Transit timing variations estimated from a collection of
  light curves} \label{sec:coll}

We present here an illustrative calculation that is relevant to the
goal of detecting planets or satellites through the perturbations they
produce on the sequence of midtransit times of a known transiting
planet. Typically an observer would fit the midtransit times
$t_{c,i}$, to a model in which the transits are strictly periodic:
\begin{eqnarray}
	t_{c,i} & = & t_{c, 0}+E_i P  \label{eq:lin}
\end{eqnarray}
for some integers $E_i$ and constants $t_{c,0}$ and $P$. Then, the
residuals would be computed by subtracting the best-fit model from the
data, and a test for anomalies would be performed by assessing the
likelihood of obtaining those residuals if the linear model were
correct. Assuming there are $N$ data points with normally-distributed,
independent errors, the likelihood is given by a
$\chi^2$-distribution, {\em prob}$(\chi^2$, $N_{\rm dof})$, where
\begin{eqnarray}
	\chi^2 = \sum_i \left[ \frac{ t_{c,i} - (t_{c,0} + E_i P)}{\sigma_{t_{c,i}}} \right]^2
\end{eqnarray}
and $N_{\rm dof} = N-2$ is the number of degrees of freedom. Values of
$\chi^2$ with a low probability of occurrence indicate the linear
model is deficient, that there are significant anomalies in the timing
data, and that further observations are warranted.

\begin{figure}[b] 
   \epsscale{0.85}
    \plotone{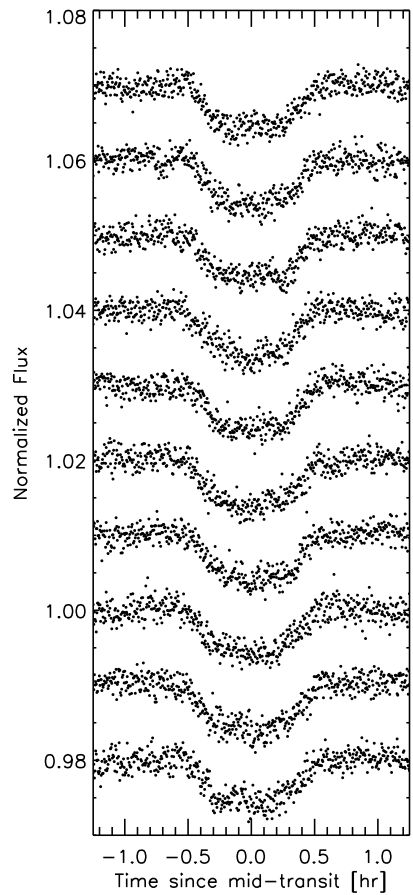}
    \caption{Simulated transit observations of the ``Hot Neptune''
      GJ~436. Arbitrary vertical offsets have been applied to the
      light curves, to separate them on the page.}
    \label{fig:coll}
 \end{figure}

We produced 10 simulated light curves of transits of the particular
planet GJ~436b, a Neptune-sized planet transiting an M dwarf (Butler
et al.~2004, Gillon et al.~2007) which has been the subject of several
transit-timing studies (see, e.g., Ribas et al.~2008, Alonso et
al.~2008, Coughlin et al.~2008). Our chosen parameters were
$R_p/R_\star = 0.084$, $a/R_\star = 12.25$, $i = 85.94$~deg, and $P =
2.644$~d. This gives $\delta=0.007$, $T=1$~hr, and $\tau=0.24$~hr.  We
chose limb-darkening parameters as appropriate for the SDSS $r$ band
(Claret 2004). We assumed that 10 consecutive transits were observed,
in each case giving $512$ uniformly-sampled flux measurements over
$2.5$~hours centered on the transit time. Noise was synthesized in the
Fourier domain (as in \S~\ref{sec:unknown}), with a white component
$\sigma_w=0.001$ and a $1/f$ component with rms $0.0005$
($\alpha=1/2$). The 10 simulated light curves are plotted in
Fig.~(\ref{fig:coll}). Visually, they resemble the best light curves
that have been obtained for this system.

\begin{figure}[htbp] 
   \epsscale{1.0}
    \plotone{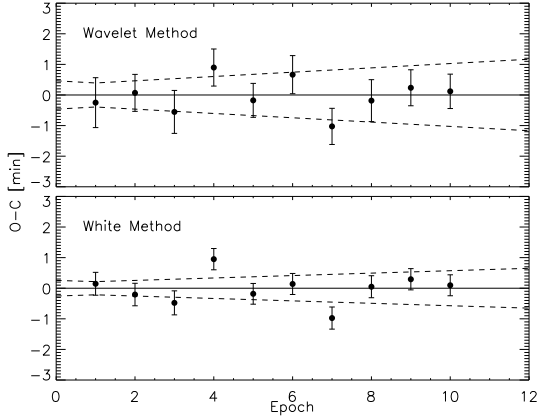}
    \caption{Transit timing variations estimated from simulated
      transit observations of GJ~436b. Each panel shows the residuals
      (observed~$-$~calculated) of a linear fit to the estimated
      midtransit times.  The midtransit times were estimated with a
      wavelet analysis and also with a white analysis, as described in
      the text. The dashed lines indicate the $1~\sigma$ errors in the
      linear model.}
    \label{fig:omc}
\end{figure}

To estimate the midtransit time of each simulated light curve, we
performed a wavelet analysis and a white analysis, allowing only the
midtransit time and the noise parameters to vary while fixing the
other parameter values at their true values. We used the same MCMC
technique that was described in \S~\ref{sec:unknown}.  Each analysis
method produced a collection of 10 midtransit times and error bars.
These 10 data points were then fitted to the linear model of
Eqn.~(\ref{eq:lin}).  Fig.~(\ref{fig:omc}) shows the residuals of the
linear fit (observed~$-$~calculated). Table~\ref{tab:omc} gives the
best-fit period for each analysis (wavelet or white), along with the
associated values of $\chi^2$.

As was expected from the results of \S~\ref{sec:unknown}, the white
analysis gave error bars that are too small, particularly for epochs 4
and 7. As a result, the practitioner of the white analysis would have
rejected the hypothesis of a constant orbital period with $98\%$
confidence. In addition, the white analysis gave an estimate for the
orbital period that is more than 1$\sigma$ away from the true value,
which might have complicated the planning and execution of future
observations. The wavelet method, in contrast, neither underestimated
nor overestimated the errors, giving $\chi^2 \approx N_{\rm dof}$ in
excellent agreement with the hypothesis of a constant orbital period.
The wavelet method also gave an estimate for the orbital period within
1$\sigma$ of the true value.

  \begin{deluxetable}{lccc}
\tablecolumns{6}
\tablewidth{0pt}
\tablecaption{Linear fits to estimated midtransit times}
\tablehead{\colhead{Method} &  \colhead{Fitted Period / True Period}  & \colhead{$\hat{\chi}^2/N_{\rm dof}$} & \colhead{{\em prob}$(\chi^2 < \hat{\chi}^2)$}}
\startdata
White  &	$1.00000071\pm0.00000043$  & 2.25 & $98\%$ \nl
Wavelet &	$1.00000048\pm0.00000077$  & 0.93 & $51\%$ 
\enddata
\label{tab:omc}
\end{deluxetable}
\newpage
\subsection{Estimation of multiple parameters} \label{sec:single}
  
Thus far we have focused exclusively on the determination of the
midtransit time, in the interest of simplicity. However, there is no
obstacle to using the wavelet method to estimate multiple parameters,
even when there are strong degeneracies among them. In this section we
test and illustrate the ability of the wavelet method to solve for all
the parameters of a transit light curve, along with the noise
parameters.
  
\begin{figure}[htbp] 
   \epsscale{1.0}
    \plotone{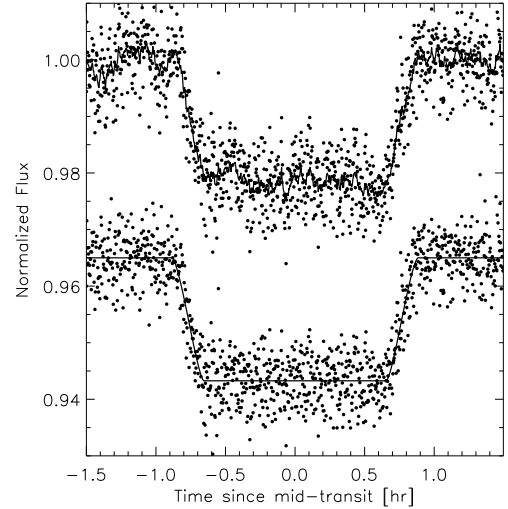}
    \caption{Wavelet analysis of a single simulated transit light
      curve.  {\it Top.}---Simulated light curve with correlated
      noise. The jagged line is the best-fitting transit model plus
      the best-fitting model of the $1/f$ component of the noise. {\it
        Bottom.}---Simulated light curve after applying the whitening
      filter of Eqn.~(\ref{eq:filter}), using the noise parameters
      estimated from the wavelet analysis.  The solid line is the
      best-fitting transit model.}
    \label{fig:test_lc}
\end{figure}

We modeled the transit as in \S\S~\ref{sec:known} and
\ref{sec:unknown}.  The noise was synthesized in the frequency domain
(as in \S~\ref{sec:unknown}), using $\sigma_w = 0.0045$, $\gamma=1$,
and $\alpha = 1/2$. The resulting simulated light curve is the upper
time series in Fig.~\ref{fig:test_lc}. We used the MCMC method to
estimate the transit parameters $\{ R_p/R_\star, a/R_\star, i, t_c\}$
and the noise parameters $\{\sigma_r, \sigma_w\}$ (again fixing
$\gamma = 1$ for simplicity).  The likelihood was evaluated with
either the wavelet method [Eqn.~(\ref{eq:lwave})] or the white method
[Eqn.~(\ref{eq:lwhite})].

\begin{figure*}[htbp] 
   \epsscale{0.8}
    \plotone{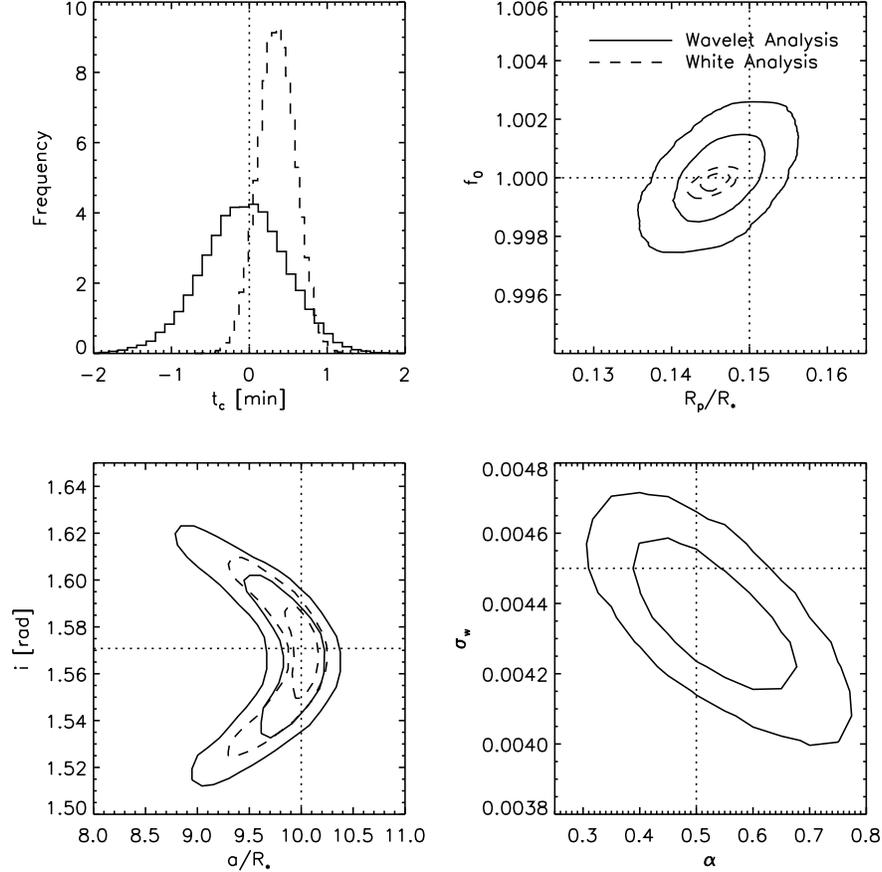}
    \caption{Results of parameter estimation for the simulated light
      curve of Fig.~\ref{fig:test_lc}.  Results for both the wavelet
      method (solid lines) and the white method (dashed lines) are
      compared.  The upper left panel shows the posterior distribution
      for the midtransit time. The other panels show confidence
      contours (68.3\% and 95.4\%) of the joint posterior distribution
      of two parameters.  The true parameter values are indicated by
      dotted lines.}
    \label{fig:test_joints}
\end{figure*}

Fig.~\ref{fig:test_joints} displays the results of this analysis in
the form of the posterior distribution for the case of $t_c$, and the
joint posterior confidence regions for the other cases. The wavelet
method gives larger (and more appropriate) confidence regions than the
white analysis. In accordance with our previous findings, the white
analysis underestimates the error in $t_c$ and gives an estimate of
$t_c$ that differs from the true value by more than 1$\sigma$. The
wavelet method gives better agreement. Both analyses give an estimate
for $R_p/R_\star$ that is smaller than the true value of $0.15$, but
in the case of the white analysis, this shift is deemed significant,
thereby ruling out the correct answer with more than 95\%
confidence. In the wavelet analysis, the true value of $R_p/R_\star$
is well within the 68\% confidence region.  Both the wavelet and white
analyses give accurate values of $a/R_\star$ and the inclination, and
the wavelet method reports larger errors. As shown in
Fig.~(\ref{fig:test_joints}), the wavelet method was successful at
identifying the parameters ($\alpha$ and $\sigma_w$) of the underlying
$1/f$ noise process.
  
\begin{figure}[htbp] 
   \epsscale{1.0}
    \plotone{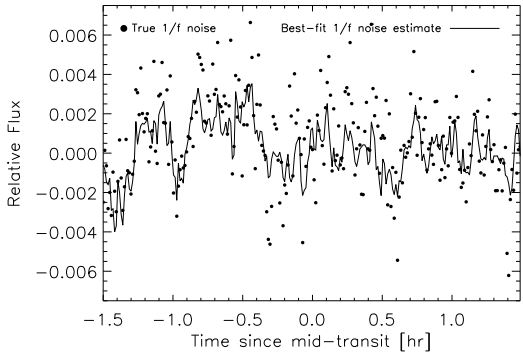}
    \caption{Isolating the correlated component. Plotted are the
      actual and estimated $1/f$ components of the noise in the
      simulated light curve plotted in Fig.~(\ref{fig:test_lc}). The
      estimated $1/f$ signal was found by applying the wavelet filter,
      Eqn.~(\ref{eq:filter}), to the residuals.}
    \label{fig:test_corr}
\end{figure}

Fig.~\ref{fig:test_lc} shows the best-fitting transit model, and also
illustrates the action of the ``whitening'' filter that was described
in \S~\ref{sec:whitening}. The jagged line plotted over the upper time
series is the best estimate of the $1/f$ contribution to the noise,
found by applying the whitening filter [Eqn.~(\ref{eq:filter})] to the
data using the estimated noise parameters. The lower time series is
the whitened data, in which the $1/f$ component has been
subtracted. Finally, in Fig.~\ref{fig:test_corr} we compare the
estimated $1/f$ noise component with the actual $1/f$ component used
to generate the data. Possibly, by isolating the correlated component
in this way, and investigating its relation to other observable
parameters, the physical origin of the noise could be identified and
understood.

\section{Summary and Discussion} \label{sec:conc}
  
In this paper we have introduced a technique for parameter estimation
based on fitting a parametric model to a time series that may be
contaminated by temporally correlated noise with a $1/f^\gamma$ power
spectral density. The essence of the technique is to calculate the
likelihood function in a wavelet basis.  This is advantageous because
a broad class of realistic noise processes produce a nearly diagonal
covariance matrix in the wavelet basis, and because fast methods for
computing wavelet transforms are available. We have tested and
illustrated this technique, and compared it to other techniques, using
numerical experiments involving simulated photometric observations of
exoplanetary transits.

For convenience we summarize the likelihood calculation here:
\begin{itemize}

\item Given the $N$ data points $y(t_i)$ obtained at evenly-spaced
  times $t_i$, subtract the model $f_i(t_i; \vec{p})$ with model
  parameters $\vec{p}$ to form the $N$ residuals $r(t_i) \equiv
  y(t_i)-f(t_i;\vec{p})$.

\item If $N$ is not a multiple of a power of two, either truncate the
  time series or enlarge it by padding it with zeros, until $N = n_0
  2^M$ for some $n_0 > 0$, $M > 0$.

\item Apply the Fast Wavelet Transform (FWT) to the residuals to
  obtain $n_0 (2^M-1)$ wavelet coefficients $r_n^m$ and $n_0$ scaling
  coefficients $\bar{r}_n^1$.

\item For stationary, Gaussian noise built from an additive
  combination of uncorrelated and correlated noise (with Power
  Spectral Density ${\cal S}(f) \propto 1/f^\gamma$), the likelihood
  for the residuals $r(t_i)$ is given by
  \begin{eqnarray}
	{\cal L} &=& \left\{ \prod_{m=2}^M \prod_{n=1}^{n_0 2^{m-1}} \frac{1}{\sqrt{2 \pi \sigma_W^2} }\exp\left[ -\frac{\left(r_n^m \right)^2}{2 \sigma_W^2}\right]  \right\}\nonumber \nopagebreak \\
	&\times&\left\{ \prod_{n=1}^{n_0}  \frac{1}{\sqrt{2 \pi \sigma_S^2} }\exp\left[ -\frac{\left(\bar{r}_n^1\right)^2}{2 \sigma_S^2}\right]\right\}
\end{eqnarray}
where 
\begin{eqnarray}
	\sigma_W^2 & = & \SIGR 2^{-\gamma m}+\sigma_w^2 \\
	\sigma_S^2 & = & \SIGR 2^{-\gamma} g(\gamma)+\sigma_w^2
\end{eqnarray}
for some noise parameters $\sigma_w > 0$, $\sigma_r > 0$ and
$g(\gamma) = O(1)$ [e.g., $g(1) \approx 0.72$].
\end{itemize}
The calculation entails the multiplication of $N$ terms and has an
overall time-complexity of $O(N)$.  With this prescription for the
likelihood function, the parameters may be optimized using any number
of traditional algorithms. For example, the likelihood may be used in
the jump-transition probability in a Monte Carlo Markov Chain
analysis, as we have done in this work.  

Among the premises of this technique are that the correlations among
the wavelet and scaling coefficients are small enough to be
negligible. In fact, the magnitude of the correlations at different
scales and times are dependent on the choice of wavelet basis and the
spectral index $\gamma$ describing the power spectral density of the
correlated component of the noise. We have chosen for our experiments
the Daubechies 4th-order wavelet basis which seems well-suited to the
cases we considered. A perhaps more serious limitation is that the
noise should be stationary. Real noise is often nonstationary. For
example, photometric observations are noisier during periods of poor
weather, and even in good conditions there may be more noise at the
beginning or end of the night when the target is observed through the
largest airmass. It is possible that this limitation could be overcome
with more elaborate noise models, or by analyzing the time series in
separate segments; future work on these topics may be warranted.

Apart from the utility of the wavelet method, we draw the following
conclusions based on the numerical experiments of \S~3. First, any
analysis that ignores possible correlated errors (a ``white'' analysis
in our terminology) is suspect, and any 2--3$\sigma$ results from such
an analysis should be regarded as provisional at best. As shown in
\S\S~\ref{sec:known}, \ref{sec:unknown}, and \ref{sec:coll}, even data
that appear ``good'' on visual inspection and that are dominated by
uncorrelated noise may give parameter errors that are underestimated
by a factor of 2--3 in a white analysis. Second, using any of the
methods described in \ref{sec:comparison} (the wavelet method, the
time-averaging method, or the residual-permutation method) is
preferable to ignoring correlated noise altogether.

Throughout this work our main application has been estimation of the
parameters of a single time series or a few such time series,
especially determining the midtransit times of transit light curves.
One potentially important application that we have not discussed is
the {\it detection}\, of transits in a database of time-series
photometry of many stars. Photometric surveys such as the ground-based
HAT (Bakos et al.~2007) and SuperWASP (Pollacco et al.~2006), and
space-based missions such as Corot (Baglin et al.~2003) and Kepler
(Borucki et al.~2003) produce tens to hundreds of thousands of time
series, spanning much longer intervals than the transit durations. It
seems likely that the parameters of a noise model could be very well
constrained using these vast databases, and that the application of a
wavelet-based whitening filter could facilitate the detection of
transits and the elimination of statistical false positives. Popular
techniques for dealing with correlated noise in large photometric
databases are those of Tamuz et al.~(2005), Kov{\'a}cs et al.~(2005),
and Pont et al.~(2006). A priority for future work is to compare these
methods with a wavelet-based method, by experimenting with realistic
survey data.

\acknowledgements We are grateful to Frederic Pont and Eric Feigelson for very detailed
and constructive critiques of an early version of this manuscript. We
also thank Scott Gaudi and Jason Eastman for helpful comments. This
work was partly supported by the NASA Origins program (grant no.\
NNX09AB33G).

\end{document}